\documentclass[11pt]{article}
\usepackage{amsmath,amsthm,bm, amscd}
\usepackage{fullpage}
\usepackage[nofullpage,full,titlepage,nousetoc,nouselot,nouselof,hylinks,final]{boaz}

\usepackage{color}

\usepackage{graphicx}

\newcommand{\supp}{\mathsf{supp}}

\newcommand{\bext}{\mathsf{BExt}}
\newcommand{\iext}{\mathsf{IExt}}

\newcommand{\zo}{\bits}

\newcommand{\srext}{\mathsf{SRExt}}
\newcommand{\baext}{\mathsf{BasicExt}}
\newcommand{\sr}{\mathsf{SR}}
\newcommand{\ssr}{\mathsf{SSR}}
\newcommand{\ind}{\mathsf{Ind}}

\def\calX{{\mathcal X}}
\def\calY{{\mathcal Y}}

\DeclareMathOperator{\expect}{E}

\newcommand{\laext}{\mathsf{laExt}}

\theoremstyle{definition}

\newcommand{\eps}{\epsilon}

\newcommand{\Supp}{\mathsf{Supp}}

  \providecommand{\stat}[1]{|#1|}

\newcommand{\Ext}{\mathsf{Ext}}

\newcommand{\cb}{\mathsf{C}}

\newcommand{\BI}{\begin{itemize}}
\newcommand{\EI}{\end{itemize}}
\newcommand{\BE}{\begin{enumerate}}
\newcommand{\EE}{\end{enumerate}}

\newtheorem{thm}{Theorem}      
\newcommand{\BT}{\begin{theorem}}   \newcommand{\ET}{\end{theorem}}
\newcommand{\BD}{\begin{definition}}   \newcommand{\ED}{\end{definition}}
\newcommand{\BCR}{\begin{corollary}} \newcommand{\ECR}{\end{corollary}}
\newtheorem{constr}[thm]{Construction}
\newcommand{\BCT}{\begin{constr}} \newcommand{\ECT}{\end{constr}}
\newcommand{\BL}{\begin{lemma}}   \newcommand{\EL}{\end{lemma}}

\newcommand{\BP}{\begin{proposition}}   \newcommand{\EP}{\end{proposition}}
\newcommand{\BCM}{\begin{claim}}   \newcommand{\ECM}{\end{claim}}
\newcommand{\BF}{\begin{fact}}   \newcommand{\EF}{\end{fact}}
\newcommand{\BA}{\begin{assumption}}   \newcommand{\EA}{\end{assumption}}
\newcommand{\tabincell}[2]{\begin{tabular}{@{}#1@{}}#2\end{tabular}}

\def\eps{\varepsilon}

\makeatletter
\def\ExtendSymbol#1#2#3#4#5{\ext@arrow 0099{\arrowfill@#1#2#3}{#4}{#5}}
\def\RightExtendSymbol#1#2#3#4#5{\ext@arrow 0359{\arrowfill@#1#2#3}{#4}{#5}}
\def\LeftExtendSymbol#1#2#3#4#5{\ext@arrow 6095{\arrowfill@#1#2#3}{#4}{#5}}
\makeatother
\newcommand\llrightarrow[2][]{\RightExtendSymbol{-}{-}{\rightarrow}{#1}{#2}}

\newcommand\llleftarrow[2][]{\RightExtendSymbol{\leftarrow}{-}{-}{#1}{#2}}

\newcommand{\hinf}{H_\infty}
\newcommand{\thinf}{\widetilde{H}_\infty}

\begin{document}

\begin{titlepage}
\def\thepage{}

\title{
Three-Source Extractors for Polylogarithmic Min-Entropy
}

\author{
Xin Li\\
Department of Computer Science\\
Johns Hopkins University\\
Baltimore, MD 21218, U.S.A.\\
lixints@cs.jhu.edu
}

\maketitle \thispagestyle{empty}

\begin{abstract}
We continue the study of constructing explicit extractors for independent general weak random sources.\ The ultimate goal is to give a construction that matches what is given by the probabilistic method --- an extractor for two independent $n$-bit weak random sources with min-entropy as small as $\log n+O(1)$. Previously, the best known result in the two-source case is an extractor by Bourgain \cite{Bourgain05}, which works for min-entropy $0.49n$; and the best known result in the general case is an earlier work of the author \cite{Li13b}, which gives an extractor for a constant number of independent sources with min-entropy $\polylog(n)$. However, the constant in the construction of \cite{Li13b} depends on the hidden constant in the best known seeded extractor, and can be large; moreover the error in that construction is only $1/\poly(n)$.

In this paper, we make two important improvements over the result in \cite{Li13b}. First, we construct an explicit extractor for \emph{three} independent sources on $n$ bits with min-entropy $k \geq \polylog(n)$. In fact, our extractor works for one independent source with poly-logarithmic min-entropy and another independent block source with two blocks each having poly-logarithmic min-entropy. Thus, our result is nearly optimal, and the next step would be to break the $0.49n$ barrier in two-source extractors. Second, we improve the error of the extractor from $1/\poly(n)$ to $2^{-k^{\Omega(1)}}$, which is almost optimal and crucial for cryptographic applications. Some of the techniques developed here may be of independent interests.
\end{abstract}
\end{titlepage}

\section{Introduction}
\emph{Randomness extractors} are fundamental objects in studying the role of randomness in computation. Motivated by the wide applications of randomness in computation (ranging from algorithms, distributed computing to cryptography and interactive proofs), the standard requirements that the randomness used should be uniform, and the fact that real world random sources are almost always biased and defective, randomness extractors are functions that transform imperfect random sources into nearly uniform random bits. In addition, these objects are especially useful in cryptographic applications, since there even originally uniform random secrets can be compromised as a result of side channel attacks. To formally define randomness extractors, we model imperfect randomness as an arbitrary probability distribution with a certain amount of entropy; and we use the standard min-entropy to measure the randomness in such an imperfect random source.

\begin{definition}
The \emph{min-entropy} of a random variable~$X$ is
\[ H_\infty(X)=\min_{x \in \supp(X)}\log_2(1/\Pr[X=x]).\]
For $X \in \zo^n$, we call $X$ an $(n,H_\infty(X))$-source, and we say $X$ has
\emph{entropy rate} $H_\infty(X)/n$.
\end{definition}

Ideally, one would hope to construct a deterministic extractor that works for any imperfect random source with a certain amount of min-entropy. However, it is easy to show that this is an impossible task. Thus the study of randomness extractors has taken two different approaches. 

The first is to give the extractor an additional independent uniform random string (i.e., make the extractor probabilistic). These extractors are called \emph{seeded extractors} and were introduced by Nisan and Zuckerman \cite{NisanZ96}. The formal definition is given below.

\begin{definition}(Seeded Extractor)\label{def:strongext}
A function $\Ext : \bits^n \times \bits^d \rightarrow \bits^m$ is  a \emph{$(k,\eps)$-extractor} if for every source $X$ with min-entropy $k$
and independent $Y$ which is uniform on $\zo^d$,
\[|\Ext(X, Y)-U_m | \leq \e.\]
It is a \emph{strong $(k,\eps)$-extractor} if in addition we have
\[ |(\Ext(X, Y), Y) - (U_m, Y)| \leq \e,\]
where $| \cdot |$ denotes the statistical distance.
\end{definition}

One can show that with a very small amount of additional random bits (called seed, and typically of length say $d=O(\log n)$), it is possible to construct extractors for all weak random sources. Moreover, even without the auxiliary uniform random bits, these extractors can be used in  many applications (such as simulating randomized algorithms using weak random sources) just by trying all possible values of the seed. Seeded extractors have also been found to be related to many other areas in computer science, and today we have nearly optimal constructions of such extractors (e.g., \cite{LuRVW03, GuruswamiUV09, DvirW08, DvirKSS09}).

However, seeded extractors are not enough for many other important applications, most notably the ones in distributed computing and cryptography, where the trick of trying all possible values of the seed does not work. Instead, in these applications we need extractors without the uniform random seed. These extractors are called \emph{seedless extractors}. Given that it is impossible to build extractors that use just a single weak random source, one natural alternative is to try to build extractors that use multiple independent weak random sources. Indeed, it seems reasonable to assume that we can find more than one independent weak sources in nature, such as stock market, thermal noise, computer mouse movements and so on. Such extractors are called independent source extractors. A formal definition is given below. 

\BD [Independent Source Extractor]
A function $\iext: (\bits^n)^t \to \bits^m$ is an extractor for independent $(n, k)$ sources that uses $t$ sources and outputs $m$ bits with error $\e$, if for any $t$ independent $(n, k)$ sources $X_1, X_2, \cdots, X_t$, we have

\[|\iext(X_1, X_2, \cdots, X_t)-U_m| \leq \e,\]
where $| \cdot |$ denotes the statistical distance.
\ED 

Constructing independent source extractors is a major problem in the area of \emph{pseudorandomness}, and has been studied for a long time. Indeed these extractors have been used in distributed computing and cryptography (e.g., the network extractor protocols in \cite{KalaiLRZ08, KalaiLR09}). Here, one natural goal is to construct extractors that use as few number of sources as possible.\ For example, in \cite{ChorG88}, Chor and Goldreich showed that the well known Lindsey's lemma gives an extractor for two independent $(n, k)$ sources with $k > n/2$. One can also use the probabilistic method to show that there exists a deterministic extractor for just two independent sources with logarithmic min-entropy, which is optimal since extractors for one weak source do not exist. In fact, the probabilistic method shows that with high probability a random function is such a two-source extractor. Thus, explicit constructions of independent source extractors is also closely related to the general problem of \emph{derandomization}.

Independent source extractors also have close connections to Ramsey graphs.\ For example, given any boolean function with two $n$-bit inputs, one can construct a bipartite graph with $N=2^n$ vertices on each side, such that two vertices are connected if and only if the output is $1$. If the function is a two-source extractor for $(n, k)$ sources, then the resulted bipartite graph has no bipartite clique or independent set of size $K=2^k$ (i.e., a Ramsey graph). With some extra efforts, this bipartite Ramsey graph can also be converted to a regular Ramsey graph. More generally, extractors that use a few (say a constant) number of sources give Ramsey hypergraphs. 

Finally, independent source extractors are also quite useful in constructing seedless extractors for other structured sources, because in many cases other structured sources can be reduced to independent sources. Two such examples are the constructions of extractors for affine sources in \cite{Li11a} and extractors for small space sources in \cite{KampRVZ06}.

However, despite considerable efforts spent on independent source extractors, the known constructions of two-source extractors are far from optimal. To date the best known two-source extractor due to Bourgain \cite{Bourgain05}, only works for entropy $k \geq (1/2-\delta)n$ for some small universal constant $\delta>0$. Quantitatively, this is just a slight improvement over the result by Chor and Goldreich \cite{ChorG88}. Given the difficulty of constructing better two-source extractors, researchers have turned to the alternative approach of constructing extractors that use a few more weak random sources, and ideally ones that only use a constant number of sources.

This approach has been quite fruitful, starting from the work of Barak, Impagliazzo and Wigderson \cite{BarakIW04}, who applied techniques from additive combinatorics to show how to extract from a constant number ($\poly(1/\delta))$ of independent $(n, \delta n)$ sources, for any constant $\delta>0$. Following this work, by using more involved techniques, Barak et al. \cite{BarakKSSW05}  constructed extractors for three independent $(n, \delta n)$ sources for any constant $\delta>0$. This was later improved by Raz \cite{Raz05} to given an extractor that works for three independent sources where only one is required to be an $(n, \delta n)$ source while the other two can have entropy as small as $k \geq \polylog(n)$. In the same paper Raz also gave an extractor for two independent sources where one is required to have entropy $k \geq (1/2+\delta)n$ for any constant $\delta>0$, and the other can have entropy as small as $k \geq \polylog(n)$. Most of these work use advanced techniques in additive combinatorics, such as sum-product theorems and incidence theorems. However, these results only achieve a constant number of sources if at least one source has min-entropy $\delta n$ for any constant $\delta>0$.

By using clever ideas related to somewhere random sources, Rao \cite{Rao06} and subsequently Barak et al. \cite{BarakRSW06} constructed extractors for general $(n, k)$ sources that use $O(\log n/\log k)$ independent sources.\ In particular, these results give extractors that only use a constant number of sources even if the min-entropy is $n^{\delta}$ for any constant $\delta>0$. They are thus a big improvement over previous results. Based on these techniques, in \cite{Li11b} the author gave an extractor for three independent $(n,k)$ sources with $k \geq n^{1/2+\delta}$ for any constant $\delta>0$. However, in the worst case where $k=\polylog(n)$, the number of sources required is still super-constant (i.e., $O(\log n/\log \log n)$). 

In a recent breakthrough \cite{Li13a, Li13b}, the author further exploited the properties of somewhere random sources and established a connection between extraction from such sources and the problem of leader election in distributed computing. Based on this connection, the author managed to construct the first explicit extractor that uses only a constant number of sources even if the entropy is as small as $\polylog(n)$ \cite{Li13b}. More specifically, for any constant $\eta>0$, the result gives an explicit extractor for min-entropy $k \geq \log^{2+\eta}n$ that uses $O(\frac{1}{\eta})+O(1)$ independent $(n, k)$ sources. This is the first explicit independent source extractor that comes close to optimal. 

However, the result in \cite{Li13b} still suffers from two drawbacks. First, the $O(1)$ term can be pretty large.\ This is because the construction first uses a seeded extractor to convert several independent $(n, k)$ sources into somewhere random sources (by using every possible value of the seed to extract from the source and then taking the concatenation), and then takes the XOR of these somewhere random sources to reduce the error. To ensure efficiently computability we need the seed length of the seeded extractor to be $O(\log n)$; while to ensure the number of sources needed is a constant, we need the error of the seeded extractor to be at most $1/\poly(n)$. Thus, we need an optimal (up to constant factors) seeded extractor in the case where the error $\e=1/\poly(n)$. For example, the extractor in \cite{LuRVW03} does not suffice because it is only optimal when the error $\e=\mathsf{exp}(-\log n/\log^{(c)} n)$, which is larger than any $1/\poly(n)$.

Suppose we have a seeded extractor with seed length $d=\log n+C\log(1/\e)$ for some constant $C>1$, then the above XOR step needs at least $C+1$ independent weak sources. One can show that the constant $C$ here must be at least $2$, thus even if we have truly optimal seeded extractors, this step requires at least $3$ sources. After that we need at least one extra source to convert the somewhere random source into another somewhere random source with the ``almost $h$-wise independent property" as in \cite{Li13b}, and we need at least two other sources to extract nearly uniform random bits. Therefore, even with truly optimal seeded extractors the construction in \cite{Li13b} requires at least $6$ independent sources.
 
Unfortunately, currently we do not have truly optimal seeded extractors, but rather extractors that are optimal up to constant factors. The two known constructions of such extractors are \cite{GuruswamiUV09} and \cite{DvirW08} (and the related \cite{DvirKSS09}), both of which first apply a condenser to transform the weak source into a new source with entropy rate $\alpha$ for some constant $\alpha>0$, and then apply an optimal seeded extractor for such sources. However, the seeded extractors for such sources may already have a big constant $C$ in the seed length. For example, the extractor by Zuckerman \cite{Zsamp} for such sources can be estimated to have $C \geq 30$, while a different construction in \cite{GuruswamiUV09} has even larger constant, potentially reaching $C \geq 100$. Other constructions such as the block source extractor used in \cite{DvirW08} have similar behavior. Therefore, by using these seeded extractors, the $O(1)$ term in the result of \cite{Li13b} can be pretty large (e.g., $\geq 30$).

Another drawback of the result in \cite{Li13b} is that the construction only achieves error $1/\poly(n)$. This kind of error is not enough for many cryptographic applications, where we typically need to have a negligible error (i.e., $n^{-\omega(1)}$).

\subsection{Our results}
In this paper, we further improve the results in \cite{Li13b}. We construct an explicit extractor for three independent sources on $n$ bits with min-entropy $k \geq \polylog(n)$. In fact, our extractor works for one independent source with poly-logarithmic min-entropy and another independent block source with two blocks each having poly-logarithmic min-entropy. We also improve the error of the extractor from $1/\poly(n)$ to $2^{-k^{\Omega(1)}}$. Specifically, we have the following theorem.

\BT
For all $n, k \in \N$ with $k \geq \log^{12} n$, there is an efficiently computable function $\iext: \bits^n \times \bits^{2n} \to \bits^m$ such that if $X$ is an $(n,k)$-source and $Y=(Y_1, Y_2)$ is an independent $(k, k)$ block source where each block has $n$ bits, then

\[\left |(\iext(X, Y), Y)-(U_m, Y) \right | \leq \e\]

and

\[\left |(\iext(X, Y), X)-(U_m, X) \right | \leq \e,\]
where $m=0.9k$ and $\e=2^{-k^{\Omega(1)}}$. \footnote{We can show that this error is strictly $n^{-\omega(1)}$.}
\ET

As a corollary this immediately gives the following theorem.

\BT
For all $n, k \in \N$ with $k \geq \log^{12} n$, there is an efficiently computable three-source extractor $\iext: (\bits^n)^3 \to \bits^m$ such that if $X, Y, Z$ are three independent $(n,k)$-sources, then

\[\left |\iext(X, Y, Z)-U_m \right | \leq \e,\]
where $m=0.9k$ and $\e=2^{-k^{\Omega(1)}}$. 
\ET

If the min-entropy $k$ is very close to $\log^2 n$, then we also have improved results over \cite{Li13a}. In particular, we have the following theorem.

\BT
For every constant $\eta>0$ and all $n, k \in \N$ with $k \geq \log^{2+\eta} n$, there is an efficiently computable extractor $\bext: (\bits^n)^t \times (\bits^n)^t \to \bits^m$ with $t=\lceil \frac{7}{\eta} \rceil+1$, such that if $X=(X_1, X_2, \cdots X_t), Y=(Y_1, Y_2, \cdots Y_t)$ are two independent $(k ,k, \cdots, k)$- block sources where each block has $n$ bits, then

\[\left |(\bext(X, Y), Y)-(U_m, Y) \right | \leq \e\]

and

\[\left |(\bext(X, Y), X)-(U_m, X) \right | \leq \e,\]
where $m=0.9k$ and $\e=2^{-k^{\Omega(1)}}$.
\ET

As a corollary, we immediately obtain the following theorem. 

\BT
For every constant $\eta>0$ and all $n, k \in \N$ with $k \geq \log^{2+\eta} n$, there is an efficiently computable extractor $\iext: (\bits^n)^t \to \bits^m$ with $t=\lceil \frac{14}{\eta} \rceil+2$ such that if $X_1, \cdots, X_t$ are $t$ independent $(n,k)$-sources, then

\[\left |\iext(X_1, \cdots, X_t)-U_m \right | \leq \e,\]
where $m=0.9k$ and $\e=2^{-k^{\Omega(1)}}$. 
\ET

For example, the above theorem gives an extractor for min-entropy $k =\log^3 n$ that uses 16 sources,  and an extractor for min-entropy $k=\log^4 n$ that uses 9 sources.

\begin{remark}
In all theorems, the constant $0.9$ can be replaced by any constant less than $1$.
\end{remark}

\tableref{table:result} summarizes our results compared to previous constructions of independent source extractors.

\begin{table}[ht] 
\centering 
\begin{tabular}{|c|c|c|c|c|} 
\hline Construction & Number of Sources &  Min-Entropy &  Output & Error\\ 
\hline \cite{ChorG88} & 2 & $k \geq (1/2+\delta) n$, any constant $\delta$ & $\Theta(n)$ & $2^{-\Omega(n)}$ \\
\hline \cite{BarakIW04} & $\poly(1/\delta)$ & $\delta n$, any constant $\delta$ & $\Theta(n)$ & $2^{-\Omega(n)}$ \\
\hline \cite{BarakKSSW05} & 3 & $\delta n$, any constant $\delta$ & $\Theta(1)$& $O(1)$ \\
\hline  \cite{Raz05} & 3 & \tabincell{l}{One source: $\delta n$, any constant $\delta$. Other\\ sources may have $k \geq \polylog(n)$.} & $\Theta(1)$& $O(1)$ \\
\hline  \cite{Raz05}& 2 & \tabincell{l}{One source: $(1/2+\delta) n$, any constant $\delta$. \\Other source may have $k \geq \polylog(n)$} & $\Theta(k)$& $2^{-\Omega(k)}$ \\
\hline \cite{Bourgain05} & 2 &  \tabincell{l}{$(1/2-\alpha_0) n$ for some small universal \\constant $\alpha_0>0$} & $\Theta(n)$& $2^{-\Omega(n)}$ \\
\hline \cite{Rao06} & 3 & \tabincell{l}{One source: $\delta n$, any constant $\delta$. Other\\ sources may have $k \geq \polylog(n)$.} & $\Theta(k)$ & $2^{-k^{\Omega(1)}}$ \\
\hline \cite{Rao06} & $O(\log n /\log k)$ & $k \geq \polylog(n)$ & $\Theta(k)$ & $k^{-\Omega(1)}$ \\
\hline \cite{BarakRSW06} & $O(\log n /\log k)$ & $k \geq \polylog(n)$ & $\Theta(k)$ & $2^{-k^{\Omega(1)}}$ \\
\hline \cite{Li11b} & 3 & \tabincell{l}{$k=n^{1/2+\delta}$, any constant $\delta$}  & $\Theta(k)$& $k^{-\Omega(1)}$ \\
\hline \cite{Li13a} & $O(\log (\frac{\log n}{\log k}))+O(1)$ & $k \geq \polylog(n)$ & $\Theta(k)$ & $k^{-\Omega(1)}$ \\
\hline \cite{Li13b} & \tabincell{l}{$O(\frac{1}{\eta})+O(1),$ \\ $O(1)$ can be large} & $k \geq \log^{2+\eta} n$ & $\Theta(k)$ & \tabincell{l}{$n^{-\Omega(1)}+$ \\$2^{-k^{\Omega(1)}}$}  \\
\hline This work & 3 & $k \geq \log^{12} n$ & $\Theta(k)$ &  $2^{-k^{\Omega(1)}}$  \\
\hline This work & $\lceil \frac{14}{\eta} \rceil+2$ & $k \geq \log^{2+\eta} n$ & $\Theta(k)$ &  $2^{-k^{\Omega(1)}}$  \\
 \hline
\end{tabular}
\caption{\textbf{Summary of Results on Extractors for Independent Sources.}} 
\label{table:result}
\end{table}



\section{Overview of The Constructions and Techniques}
Here we give a brief overview of our constructions and the techniques. To give a clear description of the ideas, we shall be informal and imprecise sometimes.

The high level idea of our constructions still follows the framework of \cite{Li13a, Li13b}. Thus, we first briefly review the construction in \cite{Li13b}.

\subsection{A brief review of the construction in \cite{Li13b}}
The constant-source extractor in \cite{Li13b} works by first obtaining a somewhere random source (SR-source for short), which is a random $N \times m$ matrix such that at least one row of the matrix is uniform. In addition, the SR-source has the stronger property that say $\frac{2}{3}$ of the rows are uniform, and moreover they are (almost) $h$-wise independent with $h=k^{\alpha}$ for some constant $0<\alpha<1$. Once we have this SR-source, we can use the lightest bin protocol from \cite{Feige99} to reduce the number of rows in the SR-source; while after each execution of the lightest bin protocol, we use the random strings in the output of the protocol as seeds to extract from another fresh weak source, using a strong seeded extractor. This way we can ensure that the resulted new random variable (not the strings from the original SR-source) is another SR-source that preserves the $h$-wise independent property (as long as the output length of the seeded extractor is small, say at most $k/(2h)$).  On the other hand the number of rows in this new SR-source has decreased a lot, roughly from $N$ to $N^{4/\sqrt{h}}$.

We can thus repeat this process until the number of rows in the SR-source becomes small enough, say $k^{1/3}$; and then we can take at most two other independent $(n,k)$ sources and use an extractor from \cite{BarakRSW06} to extract nearly uniform random bits. Since initially the number of rows in the SR-source is $\poly(n)$, $k \geq \polylog(n)$ and $h =k^{\alpha}$, a simple calculation shows that the number of iterations needed is a constant. In addition, the initial SR-source can also be obtained from a constant number of independent $(n, k)$ sources. Thus the total number of sources needed is a constant. However, as mentioned before, the step of obtaining the initial SR-source may require a large constant number of sources.

\subsection{The new construction}
We now describe our new construction of the three source extractor. Again, we will first obtain an SR-source such that say $\frac{2}{3}$ of the rows are uniform, and moreover they are (almost) $h$-wise independent with $h=k^{\alpha}$ for some constant $0<\alpha<1$. However, we will use just \emph{two} independent $(n, k)$ sources to achieve this. This is our major improvement over the construction in \cite{Li13b}. To explain the ideas, we will first show how to use \emph{three} independent $(n, k)$ sources to obtain the SR-source.

\subsubsection{Use three sources to obtain the $h$-wise independent SR-source}
In \cite{Li13b}, the initial SR-source with the $h$-wise independent property is obtained in two steps. First, one uses a constant number of independent $(n,k)$ sources to obtain a random variable that is \emph{statistically close} to an SR-source such that say $\frac{2}{3}$ of the rows are uniform (but without the $h$-wise independent property). Then one can use a single extra independent $(n,k)$ source to obtain a new SR-source with the $h$-wise independent property. It is the first step that uses a large number of independent sources. The reason is that if we take a seeded extractor with seed length $d=\log n+C\log (1/\e)$ for some $\e=1/\poly(n)$ and convert a weak source into a somewhere (close to) random source by trying all possible values of the seed and then concatenating the outputs, then the number of rows is $N=2^d > (1/\e)^C$. In addition, the best one can say about the close to uniform rows is that each one is $\e$-close to uniform (or even worse). Thus if we want the source to be statistically close to an SR-source such that $\frac{2}{3}N$ rows are simultaneously uniform, by the union bound we would need the error of the close to uniform rows to be smaller than $\e^C$. Thus, it takes the XOR of at least $C+1$ independent sources applied with the seeded extractor to reduce the error to this small. 

Here we take a completely different approach. Since eventually we need the error of the close to uniform rows in the source (obtained by applying a seeded extractor to an $(n, k)$ source $X$ and trying all possible values of the seed) to be small, we might as well just start with a seeded extractor with larger seed length, say $\ell=k^{\beta} \gg \log n$, where $0<\beta<1$ is another constant. Now if we use an optimal strong seeded extractor $\Ext_2$ such as that in \cite{GuruswamiUV09}, we can indeed show that the error of the close to uniform rows is $\e=2^{-\Omega(k^{\beta})}$, which is small enough. Moreover, by a standard averaging argument we can show that at least $0.9$ fraction of the rows are $\e$-close to uniform.

However, by naively doing this, we have increased the number of rows in the somewhere (close to) uniform source (which we will call $\bar{X}$) to $2^{\ell}=2^{k^{\beta}}$, which is super polynomial and also much larger than $1/\e$, so it seems that we have gained nothing. Fortunately, so far we have just used one weak source. Thus we can take another weak source and use it to \emph{sample} a subset of $\poly(n)$ rows from $\bar{X}$, and hopefully with high probability conditioned on the second source, the sampled subset of rows still contains a large fraction of close to uniform rows. If this is true then we are done, since now we only have $\poly(n)$ rows and the error of each close to uniform row is $\e=2^{-\Omega(\ell)}=2^{-\Omega(k^{\beta})} \ll 1/\poly(n)$; so we can show that this new source is $\poly(n)2^{-\Omega(k^{\beta})}=2^{-k^{\Omega(1)}}$-close to an SR-source such that say $\frac{2}{3}$ of the rows are uniform.  

Given this idea, it is straightforward to implement it. To sample from a set of elements using a weak random source, it suffices to take a seeded extractor, which is equivalent to a sampler as shown in \cite{Zsamp}. More specifically, take a seeded $(k'=k/2, \e')$ extractor $\Ext_1$ with seed length $d=O(\log n+\log (1/\e'))$ and output length $\ell=k^{\beta}<0.4k$ such as that in \cite{GuruswamiUV09}, we can view it as a bipartite graph with $2^n$ vertices on the left, $2^{\ell}$ vertices on the right, and left degree $2^{d}$. Thus each vertex on the left selects  a subset of right vertices with size $2^{d}$. Now if we associate the right vertices with the $2^{\ell}$ rows in $\bar{X}$, we can use another independent $(n, k)$ source $Y$ to sample a vertex on the left, which gives us a subset of the rows in $\bar{X}$ with size $2^{d}$.

We say a row in $\bar{X}$ is ``good" if it is $\e$-close to uniform. Thus at least $0.9$ fraction of the rows are good. A standard property of the $(k', \e')$ seeded extractor implies that the number of left vertices whose induced subset of rows in $\bar{X}$ contains less than $0.9-\e'$ fraction of good rows, is at most $2^{k'}$. Since $Y$ is an $(n, k)$ source, the probability of selecting a subset of rows which contains at least $0.9-\e'$ fraction of good rows is at least $1-2^{k'} 2^{-k}=1-2^{-k/2}$. Thus it suffices to take $\e'=1/4$ and we know that with probability at least $1-2^{-k/2}$ over $Y$, the selected subset of rows of $\bar{X}$ has at least $0.9-1/4>2/3$ fraction of good rows. Moreover, since $\e'=1/4$ we have that $d=O(\log n+\log (1/\e'))=O(\log n)$, therefore the size of the selected subset is $2^{d}=\poly(n)$.

Note that the above sampling process is equivalent to computing $\Ext_2(X, \Ext_1(Y, r_i))$ for all possible values $r_i$ of the $d$ bit seed of $\Ext_1$.\ Thus (although we are sampling from a set of super-polynomial size) this can be done in polynomial time. Hence, we have used two independent $(n, k)$ sources to obtain a new source $W$ such that with high probability, $W$ is statistically close to an SR-source which has $\frac{2}{3}$ fraction of uniform rows. We can now take another independent source $Z$ and use the method in \cite{Li13b} to get an SR-source with the $h$-wise independent property.

Furthermore, notice that by doing this we have reduced the error from $1/\poly(n)$ in \cite{Li13b} to $2^{-k^{\Omega(1)}}$. Essentially, with one source we can only obtain an SR-source with $\poly(n)$ rows such that some rows are $1/\poly(n)$-close to uniform; but with two independent sources we can obtain an SR-source with $\poly(n)$ rows such that some rows are $2^{-k^{\Omega(1)}}$ (or even $2^{-\Omega(k)}$)-close to uniform. In fact, this method is quite general and can be applied to any construction that involves reducing the error in an SR-source. For example, it can also be used to reduce the error of the extractor in \cite{Rao06} from $1/\poly(n)$ to $2^{-k^{\Omega(1)}}$. On the other hand, the method used in \cite{BarakRSW06} to reduce the error of the extractor in \cite{Rao06} cannot be directly applied to the construction in \cite{Li13b}, since the construction in \cite{Li13b} has a special structure (XORing several independent copies of SR-sources). 

\subsubsection{Use two sources to obtain the $h$-wise independent SR-source} 
We now describe how we can remove one source, and use just two independent $(n, k)$ sources to obtain the $h$-wise independent SR-source. First, We also briefly review the method to generate the $h$-wise independent SR-source in \cite{Li13b}. Given an SR-source $Y$ and an independent source $X$, we will use each row of $Y$ to do several rounds of alternating extraction (cf. \cite{DW09, Li12b, Li12c}) from $X$. More specifically, we divide the binary expression of the index of the row of $Y$ into blocks of size $\log h$, and for each block we run an alternating extraction from $X$ and pick an output indexed by that block. This output is then used to start the next round of alternating extraction. The final output will be the output of the alternating extraction in the last round, indexed by the last block of the binary expression of the index of that row (more details can be found in \cite{Li13b}). The new SR-source $Z$ will then be the concatenation of the outputs for all rows.

In each alternating extraction the seed length of the seeded extractor is chosen to be $\ell=k^{\beta}$, and one can show the following. For any subset of rows in $Y$ with size $h$, if all these rows are uniform (but they may depend on each other arbitrarily), then with probability $1-2^{-\Omega(\ell)}$ over the fixing of $Y$, the joint distribution of the corresponding rows in $Z$ is $2^{-\Omega(\ell)}$-close to uniform (i.e., $Z$ has the almost $h$-wise independent property). 

Now, going back to our new construction. We have already used two independent sources $Y$ and $X$ to obtain an SR-source $W$ with $N=\poly(n)$ rows, such that with probability $1-2^{-k/2}$ over the fixing of $Y$, there exists a large subset $T \subseteq [N]$ such that each row of $W$ with index in $T$ is $2^{-\Omega(\ell)}$-close to uniform. Moreover we will have $\Ext_2$ output $\ell$ bits so that each row in $W$ has length $\ell$. We will now take another optimal seeded extractor, and then use each row of $W$ as the seed to extract from $Y$ and output $k/2$ bits. Let the concatenation of these outputs be $\bar{Y}$. We will now think of $\bar{Y}$ as an SR-source, and $X$ as an independent source, and use the same method in \cite{Li13b} described above to obtain the new SR-source $Z$ from $\bar{Y}$ and $X$. 

We will show that with high probability over the fixing of $Y$,  the new SR-source $Z$ has the desired $h$-wise independent property. Note that with probability $1-2^{-k/2}$ over the fixing of $Y$, there exists a large subset $T \subseteq [N]$ such that each row of $W$ with index in $T$ is $2^{-\Omega(\ell)}$-close to uniform. If for every $y \in \Supp(Y)$ that makes this happen, we can show that conditioned on $Y=y$, the new source $Z$ also has the desired $h$-wise independent property in the subset $T$ of rows then we are done. However, this may not be the case. Thus, we want to subtract from $1-2^{-k/2}$ the probability mass of the ``bad" $y$'s which result in a $Z$ that does not have the $h$-wise independent property in the subset $T$ of rows. Towards this goal, we define a bad $y \in \Supp(Y)$ to be a string that satisfies the following two properties: \\

\noindent a) \textbf{Conditioned on the fixing of $Y=y$, there exists a large subset $T  \subseteq [N]$ such that each row of $W$ with index in $T$ is $2^{-\Omega(\ell)}$-close to uniform}, 

and 

\noindent b) \textbf{Conditioned on the fixing of $Y=y$, there exists a subset $S \subseteq T$ with $|S|=h$ such that the joint distribution of the rows of $Z$ with index in $S$ is $\e_1$ far from uniform}, where $\e_1$ is an error parameter to be chosen later.\\

Note that $S \subseteq T$, since $y$ satisfies condition a), we must have that conditioned on the fixing of $Y=y$, each row of $W$ with index in $S$ is $2^{-\Omega(\ell)}$-close to uniform. Therefore, for each $S \subseteq [N]$ with $|S|=h$ we now define an event $Bad_S$ to be the set of $y$'s in $\Supp(Y)$ that satisfies the following two properties:\\

\noindent c) \textbf{Conditioned on the fixing of $Y=y$, each row of $W$ with index in $S$ is $2^{-\Omega(\ell)}$-close to uniform}, 

and 

\noindent d) \textbf{Conditioned on the fixing of $Y=y$, the joint distribution of the rows of $Z$ with index in $S$ is $\e_1$ far from uniform}.\\

Thus every bad $y$ must belong to some $Bad_S$. Therefore to bound the probability mass of the bad $y$'s we only need to bound $\Pr[Bad_S]$ for every $S$ and then take a union bound. Now the crucial observation is that for any fixed subset $S$, property c) is determined by the $h$ random variables $R_i=\Ext_1(Y, r_i)$ with $i \in S$. Let $R$ be the concatenation of $\{R_i, i \in S\}$ (which is a deterministic function of $Y$), and define the event $A_S$ to be the set of $r$'s in $\Supp(R)$ that makes property c) satisfied, then we have $\Pr[Bad_S]=\sum_{r \in A_S} \Pr[R=r]\Pr[Bad_S|R=r]$.

Now another crucial observation is that the size of $R$ is small. Indeed, it is bounded by $h \ell=k^{\alpha+\beta}$. If we choose $\alpha, \beta$ to be such that $\alpha+\beta <1$, then the size of $R$ is $o(k)$ and we can argue that with probability $1-2^{-\ell}$ over the fixing of $R=r$, we have that $Y$ still has min-entropy at least $k-o(k)-\ell=k-o(k)>0.9k$. Moreover condition on the fixing of $R=r$ we have that $\{W_i, i \in S\}$ is a deterministic function of $X$, and is thus independent of $Y$. 

We now bound $\Pr[Bad_S|R=r]$ in two cases. First, if $H_{\infty}(Y|R=r) < 0.9k$, we will just use $\Pr[Bad_S|R=r] \leq 1$. By the above argument this happens with probability at most $2^{-\ell}$. We now consider the case where $H_{\infty}(Y|R=r) \geq 0.9k$. In this case, we know that for all $i \in S$, $W_i$ is $2^{-\Omega(\ell)}$-close to uniform. Thus the joint distribution of $\{W_i, i \in S\}$ is $h2^{-\Omega(\ell)}=2^{-\Omega(\ell)}$-close (since $h=k^{\alpha}$ and $\ell=k^{\beta}$) to a source with $h$ truly uniform rows. Ignoring the error for the moment, we can now say that for all $i \in S$, $|(\bar{Y}_i, W_i)-(U_{\ell}, W_i)| \leq 2^{-\Omega(\ell)}$. Thus for all $i \in S$, with probability $1-2^{-\Omega(\ell)}$ over the fixing of $W_i$, we have that $\bar{Y}_i$ is $2^{-\Omega(\ell)}$-close to uniform. This implies that with probability $1-h2^{-\Omega(\ell)}=1-2^{-\Omega(\ell)}$ over the fixing of $\{W_i, i \in S\}$, we have that the joint distribution of $\{\bar{Y}_i, i \in S\}$ is $h2^{-\Omega(\ell)}=2^{-\Omega(\ell)}$-close to a source with $h$ truly uniform rows. Moreover, notice that the size of $\{W_i, i \in S\}$ is also bounded by $h \ell=k^{\alpha+\beta}$. Thus again we can argue that with probability $1-2^{-\ell}$ over the fixing of $\{W_i, i \in S\}$, we have that $X$ still has min-entropy at least $k-o(k)-\ell>0.9k$. Altogether, this implies that with probability $1-2^{-\Omega(\ell)}-2^{-\ell}=1-2^{-\Omega(\ell)}$ over the fixing of $\{W_i, i \in S\}$, we have that the joint distribution of $\{\bar{Y}_i, i \in S\}$ is $2^{-\Omega(\ell)}$-close to a source with $h$ truly uniform rows, and $X$ still has min-entropy at least $0.9k$. In addition, after this further fixing of $\{W_i, i \in S\}$, we have that $\{\bar{Y}_i, i \in S\}$ is a deterministic function of $Y$, and is thus independent of $X$.

We can now use the same argument in \cite{Li13b} (treat $\{\bar{Y}_i, i \in S\}$ as the SR-source and $X$ as an independent weak source) to argue that with probability $1-2^{-\Omega(\ell)}$ over the fixing of  $\{\bar{Y}_i, i \in S\}$ (and thus also the fixing of $Y$, since $\{\bar{Y}_i, i \in S\}$ is now a deterministic function of $Y$), we have that the joint distribution of $\{Z_i, i \in S\}$ is $2^{-\Omega(\ell)}$-close to uniform. Now adding back all the errors, the above statement is still true (except for a slight change of constants in $\Omega(\cdot)$). Thus, if we set $\e_1$ to be some $2^{-\Omega(\ell)}$ appropriately, then we have that in this case $\Pr[Bad_S|R=r] \leq 2^{-\Omega(\ell)}$. Therefore, by combining the two cases, we get that $\Pr[Bad_S] \leq 2^{-\ell}+\Pr[A_S]2^{-\Omega(\ell)} \leq 2^{-\Omega(\ell)}$.

Now by the union bound we know the probability mass of the bad $y$'s is at most $\binom{N}{h}2^{-\Omega(\ell)} \leq N^h2^{-\Omega(\ell)}=2^{O(h\log n)-\Omega(\ell)}$. If we choose $\alpha, \beta$ such that $k^{\beta-\alpha} \geq C\log n$ for some large enough constant $C>1$, then we get that this probability mass is again $2^{-\Omega(\ell)}$. Also, by choosing the constant $C$ appropriately, this will also ensure that the error of the $h$-wise independent rows (which is $2^{-\Omega(\ell)}$) is less than $N^{-6h}$. This will be enough for the lightest bin protocol to work, as shown in \cite{Li13b}. All these requirements, as well as other requirements in obtaining the $h$-wise independent SR-source, can be satisfied as long as $k = \log^{2+\eta} n$ for any constant $\eta>0$ (see Algorithm~\ref{alg:bext}). 

Now we are done.\ Subtracting the probability mass of the bad $y$'s from $1-2^{-k/2}$, we get that with probability $1-2^{-\Omega(\ell)}$ over the fixing of $Y$, the source $Z$ has the desired $h$-wise independent property. 

\subsubsection{Achieving a three-source extractor}
Now that we have used two independent sources to obtain an SR-source with the $h$-wise independent property, we can use the rest of the construction in \cite{Li13b} to get an extractor. However, the direct use of the construction in \cite{Li13b} requires at least two more sources. This is because the lightest bin protocol requires at least one round, and at the end of that round we need to use a fresh source to get another SR-source. We then need to take another source in order to finish extraction. This will give us a four-source extractor. 

In order to save one source, we observe that if the entropy $k$ is a large enough polynomial in $\log n$, then $h=k^{\alpha}$ will also be large enough so that in just one iteration of the lightest bin protocol, the number of rows in the SR-source will decrease from $N=\poly(n)$ to say $N' \leq k^{1/3}$. We let the concatenation of these rows of $Z$ be $Z'$. Note that $Z'$ is a deterministic function of $Z$. By cutting the length of each row of $Z$ (if necessary) to say $\sqrt{k}$, we see that the size of $Z'$ is bounded by $N' \sqrt{k} \leq k^{5/6}$. At the end of the lightest bin protocol we will take a fresh weak source $Y_2$ (this is the third source) and use each row of $Z'$ to extract a string of length say $0.9k$ from $Y_2$ (by using an optimal seeded extractor). We let the concatenation of these outputs be $Y'$. The analysis in \cite{Li13b} implies that with high probability over the fixing of $Z$, the new source $Y'$ is also (close to) an SR-source (here it is not necessary to have the $h$-wise independent property). 

Note that $Y'$ is a deterministic function of $Y_2$ and $Z'$, and $Z'$ is deterministic function of $Z$. Moreover conditioned on the fixing of $Y$, we have that $Z$ is a deterministic function of $X$. Thus it is also true that with high probability over the fixing of $Z'$, the new source $Y'$ is close to an SR-source. Since the size of $Z'$ is $o(k)$, we can argue that with high probability over the fixing of $Z'$, the min-entropy of $X$ is $k-o(k) > 0.9k$. Moreover conditioned on the fixing of $(Y, Z')$, we have that $X$ and $Y'$ are independent. Note that $Y'$ is an SR-source with $k^{1/3}$ rows but each row has length $0.9k \gg k^{1/3}$, thus by using an extractor from \cite{BarakRSW06} we can extract random bits from $X$ and $Y'$ which are $2^{-k^{\Omega(1)}}$-close to uniform. This gives our three-source extractor with error $2^{-k^{\Omega(1)}}$. It turns out that it is enough to choose $k \geq \log^{12} n$ and $\alpha=1/6, \beta=1/3$ in this case. Also notice here that $Y$ and $Y_2$ need not be independent, but rather it suffices to have $(Y, Y_2)$ be a block source (since the analysis first conditions on the fixing of $Y$). Thus our construction actually gives an extractor for one $(n, k)$ source and another independent $(k, k)$-block source (see Algorithm~\ref{alg:iext}).

\subsubsection{Improving the results of \cite{Li13b} for smaller min-entropy}
Our three-source extractor requires $k \geq \log^{12} n$. However, if $k = \log^{2+\eta} n$ for some small constant $\eta>0$, then we can also get improved results by replacing the step of obtaining the $h$-wise independent SR-source in \cite{Li13b} with our new construction, which uses only two independent sources. This way we get a constant-source extractor with error $2^{-k^{\Omega(1)}}$.

Moreover, once we have this SR-source, running the lightest bin protocol actually does not need fully independent sources. For example, if $X=(X_1, \cdots, X_t)$ and $Y=(Y_1, \cdots, Y_t)$ are two independent block sources where each block has min-entropy $k$ conditioned on all previous blocks, then we can first obtain the SR-source $Z$ from $(X_1, Y_1)$. Now we know that with high probability conditioned on the fixing of $Y_1$, the source $Z$ has the desired property; moreover it is a deterministic function of $X$. Thus we can run the lightest bin protocol once and take a new block from $Y$ to obtain a new SR-source $Z_2$, which is a deterministic function of $Y$ conditioned on $Z$; we can then run the lightest bin protocol again and take a new block from $X$ to obtain a new SR-source $Z_3$, which is a deterministic function of $X$ conditioned on $Z_2$, and so on. This gives us an extractor for two independent block sources with each having a constant number of blocks (see Algorithm~\ref{alg:bext}).\\

\noindent{\bf Organization.}
The rest of the paper is organized as follows.\ We give some preliminaries in \sectionref{sec:prelim}. In \sectionref{sec:alter} we define alternating extraction, an important ingredient in our construction. We present our main construction of extractors in \sectionref{sec:ext}. Finally we conclude with some open problems in \sectionref{sec:conc}.

\section{Preliminaries} \label{sec:prelim}
We often use capital letters for random variables and corresponding small letters for their instantiations. Let $|S|$ denote the cardinality of the set~$S$. For $\ell$ a positive integer,
$U_\ell$ denotes the uniform distribution on $\zo^\ell$. When used as a component in a vector, each $U_\ell$ is assumed independent of the other components.
All logarithms are to the base 2.

\subsection{Probability distributions}
\begin{definition} [statistical distance]Let $W$ and $Z$ be two distributions on
a set $S$. Their \emph{statistical distance} (variation distance) is
\begin{align*}
\Delta(W,Z) \eqdef \max_{T \subseteq S}(|W(T) - Z(T)|) = \frac{1}{2}
\sum_{s \in S}|W(s)-Z(s)|.
\end{align*}
\end{definition}

We say $W$ is $\eps$-close to $Z$, denoted $W \approx_\eps Z$, if $\Delta(W,Z) \leq \eps$.
For a distribution $D$ on a set $S$ and a function $h:S \to T$, let $h(D)$ denote the distribution on $T$ induced by choosing $x$ according to $D$ and outputting $h(x)$.

\subsection{Somewhere Random Sources and Extractors}

\begin{definition} [Somewhere Random sources] \label{def:SR} A source $X=(X_1, \cdots, X_t)$ is $(t \times r)$
  \emph{somewhere-random} (SR-source for short) if each $X_i$ takes values in $\bits^r$ and there is an $i$ such that $X_i$ is uniformly distributed.
\end{definition}

\BD (Block Sources)
A distribution $X=X_1 \circ X_2 \circ \cdots, \circ X_t$ is called a $(k_1, k_2, \cdots, k_t)$ block source if for all $i=1, \cdots, t$, we have that for all $x_1 \in \Supp(X_1), \cdots, x_{i-1} \in \Supp(X_{i-1})$, $H_{\infty}(X_i|X_1=x_1, \cdots, X_{i-1}=x_{i-1}) \geq k_i$, i.e., each block has high min-entropy even conditioned on any fixing of the previous blocks. If $k_1=k_2 = \cdots =k_t=k$, we say that $X$ is a $k$ block source.
\ED

\subsection{Prerequisites from previous work}

For a strong seeded extractor with optimal parameters, we use the following extractor constructed in \cite{GuruswamiUV09}.

\BT [\cite{GuruswamiUV09}] \label{thm:optext} 
For every constant $\alpha>0$, and all positive integers $n,k$ and any $\e>0$, there is an explicit construction of a strong $(k,\e)$-extractor $\Ext: \bits^n \times \bits^d \to \bits^m$ with $d=O(\log n +\log (1/\e))$ and $m \geq (1-\alpha) k$. 
\ET

\begin{theorem} [\cite{BarakRSW06}] \label{thm:srgeneral}
For every $n,k(n)$ with $k > \log^2 n,$ and any constants $0<\eta<1$, $0< \gamma < 1/2$ such that $k^{1-2\gamma} \geq \log^{1.1} n$, there exist constants  $0<\alpha, \beta < 1$ and a polynomial time computable function $\baext:\{0,1\}^{n} \times \{0,1\}^{k^{\gamma +1}} \rightarrow \{0,1\}^m $ s.t. if $X$ is an $(n,k)$ source and $Y$ is a $(k^{\gamma} \times k)$ $(k-k^\beta)$-SR-source,

\[ | (Y , \baext(X,Y)) - (Y , U_m) | < \epsilon \]
and
\[ | (X , \baext(X,Y)) - (X , U_m) | < \epsilon \]

where $U_m$ is independent of $X,Y$, $m = (1-\eta)k$ and
$\epsilon = 2^{-k^{\alpha}}$.
\end{theorem}

\begin{remark}
The original version of \cite{BarakRSW06} requires $k > \log^{10} n$. But this is only because the output length is $m=k-k^{\Omega(1)}$, and to achieve such output length, currently the best known seeded extractor requires seed length $d=O(\log^3 (n/\e))$. If we only need to achieve output length $m = (1-\eta)k$, then we can use a seeded extractor with seed length $d=O(\log (n/\e))$, such as \cite{GuruswamiUV09}. Then it suffices to have $k > \log^2 n$ for some properly chosen $\alpha, \beta$.
\end{remark}

The following standard lemma about conditional min-entropy is implicit in \cite{NisanZ96} and explicit in \cite{MW97}.

\begin{lemma}[\cite{MW97}] \label{lem:condition} 
Let $X$ and $Y$ be random variables and let ${\cal Y}$ denote the range of $Y$. Then for all $\e>0$, one has
\[\Pr_Y \left [ H_{\infty}(X|Y=y) \geq H_{\infty}(X)-\log|{\cal Y}|-\log \left( \frac{1}{\e} \right )\right ] \geq 1-\e.\]
\end{lemma}

We also need the following lemma.

\BL [\cite{Li12c}] \label{lem:jerror}
Let $(X, Y)$ be a joint distribution such that $X$ has range $\calX$ and $Y$ has range $\calY$. Assume that there is another random variable $X'$ with the same range as $X$ such that $|X-X'| = \e$. Then there exists a joint distribution $(X', Y)$ such that $|(X, Y)-(X', Y)| = \e$.
\EL

\section{Alternating Extraction}\label{sec:alter}
As in \cite{Li13b}, an important ingredient in the construction of our extractors is the following alternating extraction protocol.

\begin{figure}[htb]
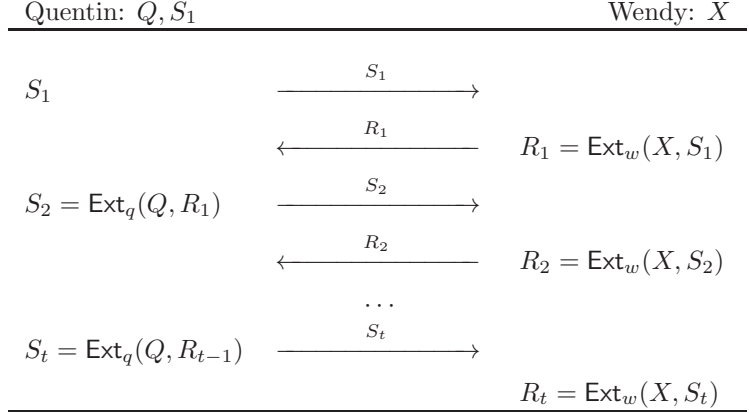

\begin{center}
\begin{small}
\begin{tabular}{l c l}
Quentin:  $Q, S_1$ & &~~~~~~~~~~Wendy: $X$ \\

\hline\\
$S_1$ & $\llrightarrow[\rule{2.5cm}{0cm}]{S_1}{} $ & \\
& $\llleftarrow[\rule{2.5cm}{0cm}]{R_1}{} $ & $R_1=\Ext_w(X, S_1)$ \\
$S_2=\Ext_q(Q, R_1)$ & $\llrightarrow[\rule{2.5cm}{0cm}]{S_2}{} $ & \\
& $\llleftarrow[\rule{2.5cm}{0cm}]{R_2}{} $ & $R_2=\Ext_w(X, S_2)$ \\
& $\cdots$ & \\
$S_t=\Ext_q(Q, R_{t-1})$ & $\llrightarrow[\rule{2.5cm}{0cm}]{S_t}{} $ & \\
& & $R_t=\Ext_w(X, S_t)$ \\
\hline
\end{tabular}
\end{small}
\caption{\label{fig:altext}
Alternating Extraction.
}
\end{center}
\end{figure}

\textbf{Alternating Extraction.} Assume that we have two parties, Quentin and Wendy. Quentin has a source $Q$,  Wendy has a source $X$. Also assume that Quentin has a uniform random seed $S_1$ (which may be correlated with $Q$). Suppose that $(Q, S_1)$ is kept secret from Wendy and $X$ is kept secret from Quentin.  Let $\Ext_q$, $\Ext_w$ be strong seeded extractors with optimal parameters, such as that in Theorem~\ref{thm:optext}. Let $\ell$ be an integer parameter for the protocol. For some integer parameter $t>0$, the \emph{alternating extraction protocol} is an interactive process between Quentin and Wendy that runs in $t$ steps. 

In the first step, Quentin sends $S_1$ to Wendy, Wendy computes $R_1=\Ext_w(X, S_1)$. She sends $R_1$ to Quentin and Quentin computes $S_2=\Ext_q(Q, R_1)$. In this step $R_1, S_2$ each outputs $\ell$ bits. In each subsequent step $i$, Quentin sends $S_i$ to Wendy, Wendy computes $R_i=\Ext_w(X, S_i)$. She replies $R_i$ to Quentin and Quentin computes $S_{i+1}=\Ext_q(Q, R_i)$. In step $i$, $R_i, S_{i+1}$ each outputs $\ell$ bits. Therefore, this process produces the following sequence: 

\begin{align*}
S_1, R_1=\Ext_w(X, S_1), S_2=\Ext_q(Q, R_1), \cdots, S_t=\Ext_q(Q, R_{t-1}), R_t=\Ext_w(X, S_t).
\end{align*}

\textbf{Look-Ahead Extractor.} Now we can define our look-ahead extractor. Let $Y=(Q, S_1)$ be a seed, the look-ahead extractor is defined as 

\[\laext(X,  Y)=\laext(X, (Q, S_1)) \eqdef R_1, \cdots, R_t.\]

The following lemma is proved in \cite{Li13b}.

\BL \label{lem:laext}
Let $Y=(Q, S_1)$ where $Q$ is an $(n_q, k_q)$ source and $S_1$ is the uniform distribution over $\ell$ bits. Let $Y_2=(Q_2, S_{21}), \cdots, Y_h=(Q_h, S_{h1})$ be another $h-1$ random variables with the same range of $Y$ that are arbitrarily correlated to $Y$. Assume that $X$ is an $(n, k)$ source independent of $(Y, Y_2, \cdots, Y_h)$, such that $k> ht\ell+10\ell+2\log(1/\e)$ and $k_q> ht\ell+10\ell+2\log(1/\e)$. Assume that $\Ext_q$ and $\Ext_w$ are strong seeded extractors that use $\ell$ bits to extract from $(n_q, 10\ell)$ sources and $(n, 10\ell)$ sources respectively, with error $\e$ and $\ell=O(\log(max\{n_q, n\})+\log(1/\e))$. Let $(R_1, \cdots, R_t)=\laext(X, Y)$ and $(R_{i1}, \cdots, R_{it})=\laext(X, Y_i)$ for $i=2, \cdots, h$. Then for any $0 \leq j \leq t-1$, we have

\[(Y, Y_2, \cdots, Y_h,  \{R_{i1}, \cdots, R_{ij}, i=2, \cdots, h\}, R_{j+1}) \approx_{\e_1} (Y, Y_2, \cdots, Y_h,  \{R_{i1}, \cdots, R_{ij}, i=2, \cdots, h\}, U_{\ell}),\]
where $\e_1=O(t\e)$.
\EL

\section{The Extractor} \label{sec:ext}
In this section we give our main construction. We will take two parameters $0<\alpha<\beta<1$ and let $h \approx k^{\alpha}$ and $\ell=k^{\beta}$. The first step is to obtain an SR-source such that a large fraction of the rows are roughly $h$-wise independent. We have the following claim and lemma.

\BCM \label{clm:extbad}
Let $\Ext: \bits^n \times \bits^d \to \bits^m$ be a $(k, \e)$ seeded extractor. For any $T \subseteq  \bits^m$ and $\rho=|T|/ 2^m$, let $Bad_T=\{x \in \bits^n: \Pr_{r \leftarrow U_d} [\Ext(x, r) \in T] > \rho+\e\}$. Then 

\[|Bad_T| \leq 2^k. \]
\ECM

\begin{proof}
Suppose not, then there exists a $T \subseteq  \bits^m$ and $\rho=|T|/ 2^m$ such that $|Bad_T| > 2^k$. Now let $X$ be the uniform distribution over the set $Bad_T$, and we have that $X$ is an $(n, k)$ source. Let $R$ be the uniform distribution over $\bits^d$. Then for any $x \in \Supp(X),$ we have that $\Pr [\Ext(x, R) \in T] > \rho+\e$. However this implies that 

\begin{align*}
& \left |\Ext(X, R)-U_m \right | \geq \left |\Pr[\Ext(X, R) \in T]-\Pr[U_m \in T] \right | \\ = &\left |\sum_{x \in \Supp(X)}\Pr[X=x] \Pr[\Ext(x, R) \in T]-\rho \right | \\> &\left |\rho+\e-\rho \right |=\e,
\end{align*}
which contradicts the fact that $\Ext$ is a $(k, \e)$ seeded extractor.
\end{proof}


\BL \label{lem:smallerror}
Let $\Ext_1: \bits^n \times \bits^d \to \bits^m$ be a $(k_1, \e_1)$ seeded extractor, and $\Ext_2: \bits^n \times \bits^m \to \bits^{m_2}$ be a $(k_2, \e_2)$ strong seeded extractor. Let $Y$ be an $(n, 2k_1)$ source and $X$ be an independent $(n, k_2)$ source. For $i=0, 1, \cdots, 2^d-1$, let $Z_i=\Ext_2(X, \Ext_1(Y, r_i))$, where $r_i$ is the $d$ bit string of $i$'s binary expression. Then with probability $1-2^{-k_1}$ over the fixing of $Y$, there exists a subset $S \subseteq \{0, 1, \cdots, 2^d-1\}$ such that the following holds:
\begin{itemize}
\item $|S| \geq (1-\sqrt{\e_2}-\e_1)2^d$.
\item $\forall i \in S$, we have $|Z_i-U_{m_2}| \leq \sqrt{\e_2}$.
\end{itemize}
\EL

\begin{proof}
Let $R$ be a uniform random string over $\bits^{m}$. Since $\Ext_2$ is a $(k_2, \e_2)$ strong seeded extractor, we have 

\[\Pr_{r \leftarrow R} [|\Ext_2(X, r)-U_{m_2}| \geq \sqrt{\e_2}] \leq \sqrt{\e_2}.\]

Let $Bad_X =\{r \in \bits^m: |\Ext_2(X, r)-U_{m_2}| \geq \sqrt{\e_2}\}$, then $|Bad_X| \leq  \sqrt{\e_2}2^m$. Now let $R'$ be the uniform distribution over $\bits^d$ and let $Bad_Y=\{y \in \bits^n: \Pr [\Ext_1(y, R') \in Bad_X] > \sqrt{\e_2}+\e_1\}$. Then by Claim~\ref{clm:extbad} we have that

\[|Bad_Y| \leq 2^{k_1}.\]

Thus if $Y$ is an $(n, 2k_1)$ source, then $\Pr_{y \leftarrow Y} [y \in Bad_Y] \leq 2^{k_1} 2^{-2k_1}=2^{-k_1}$. When $y \notin Bad_Y$, we have that $\Pr [\Ext_1(y, R') \in Bad_X] \leq \sqrt{\e_2}+\e_1$, which implies that there exists a subset $S \subseteq \{0, 1, \cdots, 2^d-1\}$ with $|S| \geq (1-\sqrt{\e_2}-\e_1)2^d$ and $\forall i \in S$, $|Z_i-U_{m_2}|=|\Ext_2(X, \Ext_1(y, r_i))-U_{m_2}| \leq \sqrt{\e_2}$.
\end{proof}


Suppose we have an $(n, k)$ source $X$ with $k \geq \polylog(n)$ and an independent SR-source $Y=Y^1 \circ \cdots \circ Y^N$ with $N=\poly(n)$ rows and each row has $0.9k$ bits, such that a large fraction of the rows are uniform. The following algorithm from \cite{Li13b} takes $X$ and $Y$ as inputs and outputs another SR-source $Z$ such that a large fraction of the rows are roughly $h$-wise independent.

\begin{algorithmsub} 
{$\ssr(X, Y)$ \cite{Li13b}}
{$X$--- an $(n,k)$-source with $k \geq \polylog(n)$. $Y=Y^1 \circ \cdots \circ Y^N$---an SR-source with $N=\poly(n)$ rows and each row has $0.9k$ bits, independent of $X$.  }
{$Z$ --- a source that is close to an SR-source.} {
Let $0<\alpha<\beta<1$ be the two constants above. Let $\ell=k^{\beta}$. Pick an integer $h$ such that $k^{\alpha} \leq h < 2k^{\alpha}$ and $h=2^{l}$ for some integer $l>0$. 
Let $\Ext_q, \Ext_w$ be strong extractors with optimal parameters from Theorem~\ref{thm:optext}, set up to extract from $((h^2+12)\ell, 10\ell)$ sources and $(n, 10\ell)$ sources respectively, with seed length $\ell$, error $\e_2=2^{-\Omega(\ell)}$ and output length $\ell$. These will be used in $\laext$.
Let $\Ext$ be a strong extractor with optimal parameters from Theorem~\ref{thm:optext}, set up to extract from $(0.9k, 2(h^2+12)\ell)$ sources, with seed length $\ell$, error $\e_2=2^{-\Omega(\ell)}$ and output length $(h^2+12)\ell$.
} {alg:ssr}

\begin{enumerate}
\item For every $i=1, \cdots, N$, use $X$ and $Y^i$ to compute $Z^i$ as follows. 
\begin{enumerate}
\item Compute the binary expression of $i-1$, which consists of $d=\log N=O(\log n)$ bits. Divide these bits sequentially from left to right into $b=\lceil \frac{d}{l} \rceil$ blocks of size $l$ (the last block may have less than $l$ bits, then we add 0s at the end to make it $l$ bits). Now from left to right, for each block $j=1, \cdots, b$, we obtain an integer $\ind_{ij} \leq 2^{l}$ such that the binary expression of $\ind_{ij}-1$ is the same as the bits in block $j$. 

\item Let $Y^{i1}$ be the first $(h+12)\ell$ bits of $Y^i$. Set $j=1$. While $j<b$ do the following. 
\begin{enumerate}
\item Compute $(R^{ij}_1, \cdots, R^{ij}_h)=\laext(X, Y^{ij})$, where $Q=Y^{ij}$ and $S_1$ is the first $\ell$ bits of $Y^{ij}$. 
\item Compute $Y^{i(j+1)}=\Ext(Y^i, R^{ij}_{\ind_{ij}})$. 
\item Set $j=j+1$. 
\end{enumerate}

\item Finally, compute $(R^{ib}_1, \cdots, R^{ib}_h)=\laext(X, Y^{ib})$ and set $Z^i=R^{ib}_{\ind_{ib}}$.
\end{enumerate}

\item Let $Z=Z^1 \circ \cdots \circ Z^N$.

\end{enumerate}
\end{algorithmsub}

We now introduce some notation as in \cite{Li13b}. For any $i \in [N]$ and $j \in [b]$, we let $Y^{i (\leq j)}$ denote $(Y^{i1}, \cdots, Y^{ij})$, let $R^{i (\leq j)}_{\ind_{i (\leq j)}}$ denote $(R^{i1}_{\ind_{i1}}, \cdots, R^{ij}_{\ind_{ij}})$ and let $f^j(i)$ denote the integer whose binary expression is the concatenation of the binary expression of $i-1$ from block $1$ to block $j$. The following lemma is proved in \cite{Li13b}.

\BL \label{lem:main}
Assume that $k \geq 2(bh+2)(h^2+12)\ell$. Fix any $v \in [N]$ such that $Y^v$ is uniform. Let $S \subset [N]$ be any subset with $|S|=h$ and $v \in S$. For any $j \in [b]$,  define $S^j_v=\{i \in S: f^j(i) < f^j(v) \}$. Then for any $j \in [b]$, we have that

\begin{align*}
&(R^{vj}_{\ind_{vj}}, \{Y^{i (\leq j)}, i \in S\}, \{R^{i j}_{\ind_{i j}}, i \in S^j_v\}, \{R^{i(\leq j-1)}_{\ind_{i(\leq j-1)}}, i \in S\}) \\ \approx_{O(jh\e_2)} &(U_{\ell}, \{Y^{i (\leq j)}, i \in S\}, \{R^{i j}_{\ind_{i j}}, i \in S^j_v\}, \{R^{i(\leq j-1)}_{\ind_{i(\leq j-1)}}, i \in S\}).
\end{align*}

Moreover, conditioned on the fixing of $(\{Y^{i (\leq j)}, i \in S\}, \{R^{i (\leq j-1)}_{\ind_{i (\leq j-1)}}, i \in S\})$, we have that 

\begin{enumerate}
\item $X$ and $Y$ are still independent.

\item $(R^{ij}_{\ind_{ij}}, i \in S)$ are all deterministic functions of $X$.

\end{enumerate}
\EL

Now we can prove the following lemma, which is slightly stronger than a similar lemma in \cite{Li13b}.

\BL \label{lem:indep}
Assume that $k \geq 2(bh+2)(h^2+12)\ell$, $X$ is an $(n, k)$-source and $Y$ is an $N \times 0.9k$ SR-source independent of $X$, with $N=\poly(n)$ such that there exists a subset $S \subset [N]$ and for any $i \in S$, $Y^i$ is uniform. Let $Z=Z^1 \circ \cdots \circ Z^N=\ssr(X, Y)$. Then for any subset $S' \subset S$ with $|S'|=h$, we have that

\[((Z^i, i \in S'), Y) \approx_{\e} (U_{h\ell}, Y),\]
where $\e=O(bh^2 \e_2)=2^{-\Omega(\ell)}$.
\EL

\begin{proof}
We order the elements in $S'$ to be $i_1 < i_2< \cdots < i_h$. Since $S' \subset S$, for any $j \in [h]$ we have that $Y^{i_j}$ is uniform. We now apply Lemma~\ref{lem:main} to the set $S'$ with $j=b$. Note that $f^b(i)=i-1$, thus for any $v \in S'$ we have $S'^b_v=\{i \in S': i < v\}$. Also note that $Z^i=R^{ib}_{\ind_{ib}}$ for any $i \in [N]$. Thus by Lemma~\ref{lem:main}, for any $j \in [h]$ we have that

\begin{align*}
&(Z^{i_j}, Z^{i_1}, \cdots, Z^{i_{j-1}}, \{Y^{i (\leq b)}, i \in S'\},  \{R^{i(\leq b-1)}_{\ind_{i(\leq b-1)}}, i \in S'\}) \\ \approx_{O(jh\e_2)} &(U_{\ell}, Z^{i_1}, \cdots, Z^{i_{j-1}}, \{Y^{i (\leq b)}, i \in S'\},  \{R^{i(\leq b-1)}_{\ind_{i(\leq b-1)}}, i \in S'\}),
\end{align*}

where $\e_2=2^{-\Omega(\ell)}$.

Note that by Lemma~\ref{lem:main}, conditioned on the fixing of $\{Y^{i (\leq b)}, i \in S'\},  \{R^{i(\leq b-1)}_{\ind_{i(\leq b-1)}}, i \in S'\}$, we have that $X$ and $Y$ are still independent, and $(R^{ib}_{\ind_{ib}}, i \in S')=(Z^i, i \in S')$ are all deterministic functions of $X$. Thus we also have 

\begin{align*}
&(Z^{i_j}, Z^{i_1}, \cdots, Z^{i_{j-1}}, \{Y^{i (\leq b)}, i \in S'\},  \{R^{i(\leq b-1)}_{\ind_{i(\leq b-1)}}, i \in S'\}, Y) \\ \approx_{O(jh\e_2)} &(U_{\ell}, Z^{i_1}, \cdots, Z^{i_{j-1}}, \{Y^{i (\leq b)}, i \in S'\},  \{R^{i(\leq b-1)}_{\ind_{i(\leq b-1)}}, i \in S'\}, Y),
\end{align*}

and therefore (since $j \leq b$)

\[(Z^{i_j}, Z^{i_1}, \cdots, Z^{i_{j-1}}, Y) \approx_{O(bh\e_2)} (U_{\ell}, Z^{i_1}, \cdots, Z^{i_{j-1}}, Y).\] 

Note this holds for every $j$, thus by a standard hybrid argument we have that

\[(Z^{i_1}, \cdots, Z^{i_h}, Y) \approx_{\e} (U_{h\ell}, Y),\]
where $\e=O(bh^2 \e_2)=O(bh^2 2^{-\Omega(\ell)})=2^{-\Omega(\ell)}$ since $\ell=k^{\beta}$, $h < 2k^{\alpha}$ and $b< \log n=k^{O(1)}$.
\end{proof}

We can now describe the algorithm to create an SR-source such that a large fraction of the rows are roughly $h$-wise independent, from just two independent sources $X$ and $Y$.

\begin{algorithmsub}
{$\sr(X, Y)$}
{$X, Y$--- two independent $(n,2k)$-source with $k \geq \polylog(n)$.}
{$Z$ --- a source that is close to an SR-source.} {
Let $0<\alpha<\beta<1$ be the two constants defined before. Let $\ell=k^{\beta}$. 
Let $\Ext_1, \Ext_2$ be two strong seeded extractors with optimal parameters from Theorem~\ref{thm:optext}, set up to extract from $(n, k)$ sources. $\Ext_1$ has seed length $d=O(\log n)$, error $\e_1=1/4$ and output length $\ell$; $\Ext_2$ has seed length $\ell$, error $\e_2=2^{-\Omega(\ell)}$ and output length $\ell$. Let $\Ext_3$ be another strong seeded extractor with optimal parameters from Theorem~\ref{thm:optext}, set up to extract from $(n, k)$ sources, with seed length $\ell$, error $\e_2$ and output length $0.9k$ (we will choose the parameters such that $2k-(h+1)\ell \geq k$).
} {alg:sr}

\begin{enumerate}
\item Let $N=2^d=\poly(n)$. For every $i=1, \cdots, N$, let $r_i$ be the $d$ bit string which is the binary expression of $i-1$. Compute $W_i=\Ext_2(X, \Ext_1(Y, r_i))$ and $Y^i=\Ext_3(Y, W_i)$. Let $\overline{Y}=Y^1 \circ \cdots \circ Y^N$. 

\item Compute $Z=\ssr(X, \overline{Y})$ using Algorithm~\ref{alg:ssr}.
\end{enumerate}
\end{algorithmsub}

We now have the following lemma.

\BL \label{lem:sr}
Assume that $k \geq 2(bh+2)(h^2+12)\ell$. There exists a constant $C>1$ such that if $\ell \geq Ch \log n$, then with probability $1-2^{-\Omega(\ell)}$ over the fixing of $Y$, the following property is satisfied:
there exists a subset $T \subseteq [N]$ such that $|T| \geq \frac{2}{3}N$ and $\forall S \subseteq T$ with $|S|=h$, we have 
\[|(Z_i, i \in S)-U_{h\ell}| \leq 2^{-\Omega(\ell)}.\]
\EL

\begin{proof}
Let $W=W_1 \circ \cdots \circ W_N$. We first show that with high probability over the fixing of $Y$, we have that $W$ is an SR-source with a large fraction of close to uniform rows. This follows directly from Lemma~\ref{lem:smallerror}. Specifically, the lemma implies that with probability $1-2^{-k}$ over the fixing of $Y$, there exists a subset $T \subseteq [N]$ with $N=2^d=\poly(n)$ such that $|T| \geq (1-\sqrt{\e_2}-1/4)N > \frac{2}{3}N$ since $\e_2=2^{-\Omega(\ell)}$; and $\forall i \in T$, we have $|W_i-U_{\ell}| \leq \sqrt{\e_2}=2^{-\Omega(\ell)}$.

Now consider any $y \in \Supp(Y)$ which makes the above happen.\ We'd like to show that conditioned on this $Y=y$, in the final output $Z$, the same set $T$ of the rows will also have the property of being roughly $h$-wise independent.\ However, this may not be the case; and if not, we will call such a $y$ bad. Now fix any bad $y$. Then we know that there must be a subset $S \subset T$ with $|S|=h$ such that $|(Z^i, i \in S)-U_{h\ell}| > \e'$ for some $\e'=2^{-\Omega(\ell)}$. At the same time, since $S \subset T$ we also know that $\forall i \in S$, we have $|W_i-U_{\ell}| \leq \sqrt{\e_2}=2^{-\Omega(\ell)}$. Let 

\[Bad_S=\{y \in \Supp(Y): \forall i \in S, |W_i-U_{\ell}| \leq \sqrt{\e_2} \text{ but } |(Z^i, i \in S)-U_{h\ell}| >\e'\}\]
for some $\e'=2^{-\Omega(\ell)}$. Then we must have $y \in Bad_S$. Therefore, any bad $y$ must be in $\bigcup_S Bad_S$. By the union bound we know 

\[\Pr_{y \leftarrow Y} [y \text{ is bad}] \leq \sum_S \Pr[Bad_S].\]

Thus to bound the probability of a bad $y$ we only need to bound $\Pr[Bad_S]$. 

Now fix any subset $S \subseteq [N]$ with $|S|=h$. Let $R=\{\Ext_1(Y, r_i), i \in S\}$. We now bound $\Pr[Bad_S]$ as follows. Define

\[A_S=\{r \in \Supp(R): \forall i \in S, |W_i-U_{\ell}| \leq \sqrt{\e_2}\}.\]

Then 

\[\Pr[Bad_S]=\sum_{r \in A_S} \Pr[R=r]\Pr[Bad_S|R=r].\]

We now estimate $\Pr[Bad_S|R=r]$. First we know that conditioned on any $R=r$, we have that $\forall i \in S, |W_i-U_{\ell}| \leq \sqrt{\e_2}$. Thus by Lemma~\ref{lem:jerror} we can get rid of the error one by one for each $i \in S$ and we have that there exists another random variable $(W'_i, i \in S)$ such that $\forall i \in S$, $W'_i=U_{\ell}$ and $|(W_i, i \in S)-(W'_i, i \in S)| \leq h\sqrt{\e_2}$. From now on we'll think of $(W_i, i \in S)$ as being $(W'_i, i \in S)$ (i.e., every row is truly uniform). This only adds $h\sqrt{\e_2}$ to the final error. Now, since the size of $R$ is bounded by $h\ell$, by Lemma~\ref{lem:condition} we have that

\[\Pr_{r \leftarrow R} \left [ H_{\infty}(Y|R=r) \geq 2k-h\ell-\ell \geq k \right ] \geq 1-2^{-\ell}.\]

Now we have the following two cases.

\textbf{Case 1}: $H_{\infty}(Y|R=r)<k$. In this case we'll just bound $\Pr[Bad_S|R=r]$ by $\Pr[Bad_S|R=r] \leq 1$. However, the probability of such $R=r$ is at most $2^{-\ell}$.

\textbf{Case 2:} $H_{\infty}(Y|R=r) \geq k$. In this case, we know that $\forall i \in S$, $W_i$ is uniform and independent of $Y$ (since it is a deterministic function of $X$ conditioned on the fixing of $R=r$). Thus by Theorem~\ref{thm:optext} we have that 

\[|(Y^i, W_i)-(U_{0.9k}, W_i)| \leq \e_2.\]

Therefore $\forall i \in S$, we have that with probability $1-\sqrt{\e_2}$ over the fixing of $W_i$, $Y^i$ is $\sqrt{\e_2}$-close to uniform. Let $W=\{W_i, i \in S\}$. Then with probability $1-h\sqrt{\e_2}$ over the fixing of $W$, we have that each $Y^i$ is $\sqrt{\e_2}$-close to uniform. Thus again by Lemma~\ref{lem:jerror}, we have that $Y^S=\{Y^i, i \in S\}$ is $h\sqrt{\e_2}$-close to another source $Y'^S=\{Y'^i, i \in S\}$ where $\forall i, Y'^i=U_{0.9k}$. Now since the size of $W$ is $h\ell$, again by Lemma~\ref{lem:condition} we have that with probability $1-2^{-\ell}$ over the fixing of $W$, $X$ still has min-entropy at least $k$. Thus, in summary, with probability $1-h\sqrt{\e_2}-2^{-\ell}$ over the fixing of $W$, we have that $X$ has min-entropy at least $k$, $Y^S=\{Y^i, i \in S\}$ is $h\sqrt{\e_2}$-close to $Y'^S=\{Y'^i, i \in S\}$, and $X$ and $Y^S$ are independent (since $W$ is a deterministic function of $X$). Assume for now that $Y^S$ is just $Y'^S$, then we can apply Lemma~\ref{lem:indep} to conclude that in this case, we have

\[|((Z^i, i \in S), Y)-(U_{h\ell}, Y)| \leq O(bh^2\e_2).\]

Therefore with probability $1-O(bh\sqrt{\e_2})$ over the fixing of $Y$, we have that $|(Z^i, i \in S)-U_{h\ell}| \leq h\sqrt{\e_2}$. Now adding back all the errors, we get that with probability $1-O(bh\sqrt{\e_2})-h\sqrt{\e_2}=1-O(bh\sqrt{\e_2})$ over the fixing of $Y$, we have that 

\[|(Z^i, i \in S)-U_{h\ell}| \leq h\sqrt{\e_2}+h\sqrt{\e_2}+h\sqrt{\e_2}+2^{-\ell} \leq (3h+1)\sqrt{\e_2}.\]

Now let $\e'=(3h+1)\sqrt{\e_2}=2^{-\Omega(\ell)}$ since $\e_2=2^{-\Omega(\ell)}$, $\ell=k^{\beta}$ and $h<2k^{\alpha}$. We have that in Case 2, 

\[\Pr[Bad_S|R=r] \leq O(bh\sqrt{\e_2}).\]

Therefore for any fixed $S$, we have that 

\[\Pr[Bad_S] \leq 2^{-\ell}+\Pr[A_S]O(bh\sqrt{\e_2})=O(bh\sqrt{\e_2})=2^{-\Omega(\ell)},\]
since $b < \log n=k^{O(1)}$ and $h<2k^{\alpha}$. 

Thus

\[\Pr_{y \leftarrow Y} [y \text{ is bad}] \leq \binom{N}{h} 2^{-\Omega(\ell)} <N^h 2^{-\Omega(\ell)}=2^{-\Omega(\ell)+O(h \log n)}=2^{-\Omega(\ell)},\]
if we choose $h, \ell$ such that $\ell \geq Ch \log n$ for some sufficiently large constant $C>1$.

Now subtracting the probability mass of the bad $y$'s, we get that with probability $1-2^{-k}-2^{-\Omega(\ell)}=1-2^{-\Omega(\ell)}$ over the fixing of $Y$, there exists a subset $T \subseteq [N]$ such that $|T| \geq \frac{2}{3}N$ and $\forall S \subseteq T$ with $|S|=h$, we have 
\[|(Z_i, i \in S)-U_{h\ell}| \leq \e'=2^{-\Omega(\ell)}.\]
\end{proof}

Next we describe the lightest bin protocol, defined in \cite{Li13a}.

\textbf{Lightest bin protocol:}
Assume there are $N$ strings $\{z^i, i \in [N]\}$ where each $z_i \in \bits^m$ with $m > \log N$. The output of a lightest bin protocol with $r< N$ bins is a subset $T \subset [N]$ that is obtained as follows. Imagine that each string $z^i$ is associated with a player $P_i$. Now, for each $i$, $P_i$ uses the first $\log r$ bits of $z_i$ to select a bin $j$, i.e., if the first $\log r$ bits of $z_i$ is the binary expression of $j-1$, then $P_i$ selects bin $j$. Now let bin $l$ be the bin that is selected by the fewest number of players. Then 

\[T=\{i \in [N]: P_i \text{ selects bin } l.\}\]

The following lemma is proved in \cite{Li13b}.


\BL \label{lem:bin}
For every constant $0<\gamma<1$ there exists a constant $C_1>1$ such that the following holds. For any $n, k, m, N \in \N$, any even integer $h \geq C_1$ and any $\e>0$ with $N \geq h^2$, $\e < N^{-6h}$, $k > 20h(\log n+ \log (1/\e))$ and $m> 10(\log n+ \log (1/\e))$,\footnote{The constants actually depend on the hidden constant in the seed length $d=O(\log(n/\e))$ of an optimal seeded extractor. Nevertheless they are always constants and don't really affect our analysis. For simplicity and clarity we use  20, 10 here.} assume that we have $N$ sources $\{Z^i_1, i \in [N]\}$ over $m$ bits and a subset $S \subset [N]$ with $|S| \geq \delta N$ for some constant $\delta>1/2$, such that for any $S' \subset S$ with $|S'|=h$, we have

\[(Z^i_1, i \in S') \approx_{\e} U_{hm}.\]

Let $Z_1=Z^1_1 \circ \cdots \circ Z^N_1$. Use $Z_1$ to run the lightest bin protocol with $r=\frac{\gamma^2}{16h}N^{1-\frac{2}{\sqrt{h}}}$ bins \footnote{For simplicity, we assume that $r$ is a power of 2. If not, we can always replace it with a power of 2 that is at most $2r$. This does not affect our analysis.} and let the output contain $N_2$ elements $\{i_1, i_2, \cdots, i_{N_2} \in [N]\}$. Assume that $X$ is an $(n, k)$ source independent of $Z_1$. For any $j \in [N_2]$, let $Z^j_2=\Ext(X, Z^{i_j}_1)$ where $\Ext$ is the strong seeded extractor in theorem~\ref{thm:optext} that has seed length $m$ and outputs $m_2=k/(2h)$ bits with error $\e$. Then with probability at least $1-N^{-\sqrt{h}/2}$ over the fixing of $Z_1$, there exists a subset $S_2 \subset [N_2]$ with $|S_2| \geq \delta(1-\gamma)N/r \geq \delta(1-\gamma) N_2$ such that for any $S'_2 \subset S_2$ with $|S'_2|=h$, we have

\[(Z^i_2, i \in S'_2) \approx_{\e_2} U_{hm_2}\]
with $\e_2 < N^{-6h}_2$ and $m_2 > 10(\log n+ \log (1/\e_2))$.
\EL

We can now present our construction of extractors for independent sources.

\begin{algorithmsub}
{Independent Source Extractor $\iext$}
{$X$ --- an $(n,2k)$-source with $k \geq \frac{1}{2}\log^{12} n$. $Y=(Y_1, Y_2)$ --- a $(2k, 2k)$ block source where each block has $n$ bits, independent of $X$.}
{$V$ --- a random variable close to uniform.} {
Let $\sr$ be the function in Algorithm~\ref{alg:sr}. Let $\baext$ be the extractor in Theorem~\ref{thm:srgeneral}. Let $\Ext$ be the strong extractor in Theorem~\ref{thm:optext}. Let $0< \alpha<\beta<1$ be the two constants defined before. Let $0<\gamma<1$ be the constant in Lemma~\ref{lem:bin}. We will choose $\alpha=1/6, \beta=1/3$ and $\gamma=1/4$. Let $h, \ell$ be the two parameters in Algorithm~\ref{alg:ssr} with $k^{\alpha} \leq h < 2k^{\alpha}$ and $\ell=k^{\beta}$.
} {alg:iext}

\begin{enumerate}
\item Compute $Z=Z^1\circ \cdots \circ Z^N=\sr(X, Y_1)$.

\item Let $N=\poly(n)$ be the number of rows in $Z$. Run the lightest bin protocol with $Z$ and $r=\frac{\gamma^2}{16h}N^{1-\frac{2}{\sqrt{h}}}$ bins and let the output contain $N_1$ elements $\{i_1, i_2, \cdots, i_{N_1} \in [N]\}$. Let $Z_1=Z^1_1\circ \cdots \circ Z^{N_1}_1$ be the concatenation of the corresponding rows  in $Z$ (i.e., $Z^j_1=Z^{i_j}$). 

\item Note that $N_1 \leq \lfloor N/r \rfloor$. Without loss of generality assume that $N_1 = \lfloor N/r \rfloor$. If not, add rows of all $0$ strings to $Z_1$ until $N_1 = \lfloor N/r \rfloor$. 

\item For any $j \in [N_1]$, compute $Z^j_2=\Ext(Y_2, Z^j_1)$ and output $m_2=\sqrt{k}$ bits. Let $Z_2= Z^1_2 \circ \cdots \circ Z^{N_1}_2$.

\item For any $j \in [N_1]$, compute $Z^j_3=\Ext(X, Z^j_2)$ and output $m_3=1.9k$ bits. Let $Z_3= Z^1_3 \circ \cdots \circ Z^{N_1}_3$.
 
\item Compute $V=\baext(Y_2, Z_3)$.
\end{enumerate}
\end{algorithmsub}

We now have the following theorem.

\BT \label{thm:iext}
There exists a constant $C_0>1$ such that for any $n, k \in \N$ with $n \geq C_0$ and $k \geq \frac{1}{2}\log^{12} n$, if $X$ is an $(n,2k)$-source and $Y=(Y_1, Y_2)$ is an independent $(2k, 2k)$ block source where each block has $n$ bits, then

\[\left |(\iext(X, Y), Y)-(U_m, Y) \right | \leq \e\]

and

\[\left |(\iext(X, Y), X)-(U_m, X) \right | \leq \e,\]
where $m=1.8k$ and $\e=2^{-k^{\Omega(1)}}$.
\ET

\begin{thmproof}
By Lemma~\ref{lem:sr}, with probability $1-2^{-\Omega(\ell)}$ over the fixing of $Y_1$, there exists a subset $T \subseteq [N]$ such that $|T| \geq \frac{2}{3}N$ and $\forall S \subseteq T$ with $|S|=h$, we have 

\[|(Z^i, i \in S)-U_{h\ell}| \leq 2^{-\Omega(\ell)}.\]

We now want to apply Lemma~\ref{lem:bin}. But first let's check that the conditions of Lemma~\ref{lem:sr} and Lemma~\ref{lem:bin} are satisfied. Note that $k^{\alpha} \leq h < 2k^{\alpha}$, $\ell=k^{\beta}$ and $b< \log n$. To apply Lemma~\ref{lem:sr}, we need that $k \geq 2(bh+2)(h^2+12)\ell$ and $\ell \geq Ch\log n$ for some sufficiently large constant $C>1$.\ To apply Lemma~\ref{lem:bin}, we need that $\e'< N^{-6h}$, $k > 20h(\log n+ \log (1/\e'))$ and $m=\ell> 10(\log n+ \log (1/\e'))$. In Algorithm~\ref{alg:sr} we also need $k \geq (h+1)\ell$.\ Altogether, it suffices to have $0<\alpha<\beta<1$ satisfy the following conditions.


\[k \geq 3\log n h^3 \ell, \text{  } \ell \geq Ch\log n, \text{} \e' < N^{-6h} \text{ and } \ell> 10(\log n+ \log (1/\e')).\]

These conditions are satisfied if the following conditions are satisfied.

\[k \geq 24 k^{3\alpha+\beta}\log n \text{ and } \ell=k^{\beta} \geq Ck^{\alpha}\log n\]
for some constant $C>1$.

Now if $\alpha=1/6, \beta=1/3$ and $k \geq \frac{1}{2}\log^{12} n$, then we see that for sufficiently large $n$,

\[\frac{k}{k^{3\alpha+\beta}}=k^{1/6} \geq \Omega(\log^2 n) > 24 \log n \text{ and } \frac{k^{\beta}}{k^{\alpha}}=k^{1/6} \geq \Omega(\log^2 n) > C \log n.\]

Thus the above conditions are satisfied.

Notice that $m_2=\sqrt{k} < k/(2h)$, thus by Lemma~\ref{lem:bin} we have that with probability at least $1-N^{-\sqrt{h}/2}$ over the fixing of $Z$, there exists a subset $S \subset [N_1]$ with $|S| \geq \delta(1-\gamma)N/r \geq \frac{2}{3} \frac{3}{4}N/r=\frac{1}{2}N/r$ such that for any $S' \subset S$ with $|S'|=h$, we have

\[(Z^i_2, i \in S') \approx_{\e_2} U_{h\sqrt{k}}\]
with $\e_2 < N^{-6h}_1$.
 
Note that $Z_2$ is a deterministic function of $Y_2$ and $Z_1$, and $Z_1$ is a deterministic function of $Z$. Thus we also have that with probability at least $1-N^{-\sqrt{h}/2}$ over the fixing of $Z_1$, the above property holds. Also note that $N/r=\frac{16h}{\gamma^2}N^{\frac{2}{\sqrt{h}}}>16h$, so $|S| > 8h>1$. Thus with probability at least $1-N^{-\sqrt{h}/2}$ over the fixing of $Z_1$, we have that $Z_2$ is $N^{-6h}_1 <(8h)^{-6h}$-close to an SR source (since $N_1 \geq |S|$).

Note that conditioned on the fixing of $Z_1$, we have that $Z_2$ is a deterministic function of $Y_2$, and is thus independent of $X$. Now note that $N/r=\frac{16h}{\gamma^2}N^{\frac{2}{\sqrt{h}}}$. Since $h \geq k^{\alpha}=k^{1/6}$ and $k \geq \frac{1}{2}\log^{12} n$, we have that

\[N^{\frac{2}{\sqrt{h}}} \leq \poly(n)^{O(1/\log n)}=O(1).\]

Thus $N_1 \leq N/r=O(h)< k^{1/4}$. Note that conditioned on the fixing of $Y_1$, we have that $Z_1$ is a deterministic function of $X$, with the size of $Z_1$ bounded by $ k^{1/4}\ell <k^{2/3}$. Therefore by Lemma~\ref{lem:condition}, we have that with probability $1-2^{-0.05k}$ over the fixing of $Z_1$, $X$ still has min-entropy at least $2k-k^{2/3}-0.05k > 1.94k$.

Now since $Z_2$ is independent of $X$ and assuming that $Z_2$ is indeed an SR-source, then by Theorem~\ref{thm:optext} we have that for some $i \in [N_1]$,

\[|(Z^i_3, Z^i_2)-(U_{1.9k}, Z^i_2)| \leq 2^{-\Omega(\sqrt{k})}.\]

Thus with probability at least $1-2^{-\Omega(\sqrt{k})}$ over the fixing of $Z^i_2$ (and thus also the fixing of $Z_2$), we have that $Z_3$ is $2^{-\Omega(\sqrt{k})}$-close to an $N_1 \times 1.9k$ SR-source. Moreover, conditioned on the further fixing of $Z_2$, we have that $Z_3$ is a deterministic function of $X$, and is thus independent of $Y_2$. Furthermore, note the size of $Z_2$ is bounded by $N_1 \sqrt{k} \leq k^{1/4}\sqrt{k}=k^{3/4}$. Thus again by Lemma~\ref{lem:condition}, we have that with probability $1-2^{-0.05k}$ over the fixing of $Z_2$, $Y_2$ still has min-entropy at least $2k-k^{3/4}-0.05k > 1.94k$.

Note that $N_1< k^{1/4}$ and $k^{1-2/4}=k^{1/2}> \log^{1.1} n$, thus by Theorem~\ref{thm:srgeneral}, we have that 

\[\left |(V, Y_2)-(U_m, Y_2) \right | \leq \e_2\]

and

\[\left |(V, Z_3)-(U_m, Z_3) \right | \leq \e_2,\]
where $m=1.8k$ and $\e_2=2^{-k^{\Omega(1)}}$. Since we have already fixed $Y_1$, $Z_1$ and $Z_2$, we have that $Z_3$ is a deterministic function of $X$. Thus conditioned on $Z_3$, we have that $V$ is a deterministic function of $Y_2$, which is independent of $X$. Thus we also have that 

\[\left |(V, X)-(U_m, X) \right | \leq \e\]

and

\[\left |(V, Y)-(U_m, Y) \right | \leq \e,\]

where by adding back all the errors we have 

\[\e \leq \e_2+2^{-\Omega(\ell)}+N^{-\sqrt{h}/2}+(8h)^{-6h}+2^{-0.05k}+2^{-\Omega(\sqrt{k})}+2^{-\Omega(\sqrt{k})}+2^{-0.05k}=2^{-k^{\Omega(1)}}. \footnote{One can show that in this case $\e_2$ is $n^{-\omega(1)}$, as well as all the other terms. So the entire error is $n^{-\omega(1)}$.}
\]
\end{thmproof}

Note that when $n < C_0$, the extractor can be constructed in constant time just by exhaustive search (in fact, we can get a two-source extractor in this way). Thus, we have the following theorem (by replacing $2k$ with $k)$.

\BT
For all $n, k \in \N$ with $k \geq \log^{12} n$, there is an efficiently computable function $\iext: \bits^n \times \bits^{2n} \to \bits^m$ such that if $X$ is an $(n,k)$-source and $Y=(Y_1, Y_2)$ is an independent $(k, k)$ block source where each block has $n$ bits, then

\[\left |(\iext(X, Y), Y)-(U_m, Y) \right | \leq \e\]

and

\[\left |(\iext(X, Y), X)-(U_m, X) \right | \leq \e,\]
where $m=0.9k$ and $\e=2^{-k^{\Omega(1)}}$. \footnote{The constant $0.9$ can be replaced by any constant less than $1$.}
\ET

As a corollary, we immediately obtain the following theorem. 

\BT
For all $n, k \in \N$ with $k \geq \log^{12} n$, there is an efficiently computable three-source extractor $\iext: (\bits^n)^3 \to \bits^m$ such that if $X, Y, Z$ are three independent $(n,k)$-sources, then

\[\left |\iext(X, Y, Z)-U_m \right | \leq \e,\]
where $m=0.9k$ and $\e=2^{-k^{\Omega(1)}}$. 
\ET

If the entropy $k$ gets very close to $\log^2 n$, then we can use a similar construction as the extractor in \cite{Li13b}, except replacing the step of creating the initial SR-source with the method in this paper. In this case we can get an extractor for two independent block sources each with a constant number of blocks of min-entropy $k$. We have the following algorithm.

\begin{algorithmsub}
{Block Source Extractor $\bext$}
{$X=(X_1, X_2, \cdots X_t), Y=(Y_1, Y_2, \cdots Y_t)$ --- two independent $(2k ,2k, \cdots, 2k)$-block sources where each block has $n$ bits and $k \geq \frac{1}{2}\log^{2+\eta} n$ for any constant $\eta>0$. }
{$W$ --- a random variable close to uniform.} {
Let $\sr$ be the function in Algorithm~\ref{alg:sr}. Let $\baext$ be the extractor in Theorem~\ref{thm:srgeneral}. Let $\Ext$ be the strong extractor in Theorem~\ref{thm:optext}. Let $\alpha=\frac{\mu}{6(2+\mu)}$ and $\beta=\frac{6+2\mu}{6(2+\mu)}$, where $\mu=0.95\eta$ be the two constants defined before, and $\gamma=\frac{\eta}{70}$ be another constant.  Let $h, \ell$ be the two parameters in Algorithm~\ref{alg:ssr} with $k^{\alpha} \leq h < 2k^{\alpha}$ and $\ell=k^{\beta}$.
} {alg:bext}

\begin{enumerate}
\item Compute $Z_1=Z^1_1\circ \cdots \circ Z^{N_1}_1=\sr(X_1, Y_1)$ where $N_1=\poly(n)$. Set the boolean indicator $v_y=1$.

\item Set $t=1$. While $N_t$ (the number of rows in $Z_t$) is bigger than $\frac{16h^3}{\gamma^2}$ do the following:

\begin{enumerate}
\item Run the lightest bin protocol with $Z_t$ and $r_t=\frac{\gamma^2}{16h}N_t^{1-\frac{2}{\sqrt{h}}}$ bins and let the output contain $N_{t+1}$ elements $\{i_1, i_2, \cdots, i_{N_{t+1}} \in [N_t]\}$. 

\item If $v_y=1$, take a fresh new block $Y'$ from $Y$, and for any $j \in [N_{t+1}]$, compute $Z^j_{t+1}=\Ext(Y', Z^{i_j}_t)$ and output $\ell \leq k/(2h)$ bits (note that we have $k \geq 2h \ell$ by our choices of $\alpha, \beta$). Set  $v_y=0$. Otherwise, take a fresh new block $X'$ from $X$, and for any $j \in [N_{t+1}]$, compute $Z^j_{t+1}=\Ext(X', Z^{i_j}_t)$ and output $\ell \leq k/(2h)$ bits. Set  $v_y=1$.
 
\item Let $Z_{t+1}=Z^1_{t+1} \circ \cdots \circ Z^{N_{t+1}}_{t+1}$. Set $t=t+1$.
\end{enumerate}

\item At the end of the above iteration we get a source $Z_t$ with $N_t \leq \frac{16h^3}{\gamma^2}$ rows. Without loss of generality assume that at this time $v_y=0$ (otherwise switch the roles of $X$ and $Y$), and the last two blocks of $X, Y$ used are $X', Y'$. For any $j \in [N_t]$, compute $Z'^j=\Ext(X', Z^j_t)$ and output $m_2=1.9k$ bits. Let $Z'= Z'^1 \circ \cdots \circ Z'^{N_t}$.

\item Compute $W=\baext(Y', Z')$.
\end{enumerate}

\end{algorithmsub}

We now have the following theorem.

\BT
For every constant $\eta>0$ there exists a constant $C_0>1$ such that for any $n, k \in \N$ with $n \geq C_0$ and $k \geq \frac{1}{2}\log^{2+\eta} n$, if $X=(X_1, X_2, \cdots X_t), Y=(Y_1, Y_2, \cdots Y_t)$ are two independent $(2k ,2k, \cdots, 2k)$-block sources where each block has $n$ bits and $t=\lceil \frac{7}{\eta} \rceil+1$, then

\[\left |(\bext(X, Y), Y)-(U_m, Y) \right | \leq \e\]

and

\[\left |(\bext(X, Y), X)-(U_m, X) \right | \leq \e,\]
where $m=1.8k$ and $\e=2^{-k^{\Omega(1)}}$.
\ET

\begin{thmproof}(Sketch)
By Lemma~\ref{lem:sr}, with probability $1-2^{-\Omega(\ell)}$ over the fixing of $Y_1$, there exists a subset $T \subseteq [N]$ such that $|T| \geq \frac{2}{3}N$ and $\forall S \subseteq T$ with $|S|=h$, we have 

\[|(Z^i_1, i \in S)-U_{h\ell}| \leq 2^{-\Omega(\ell)}\].

We now want to apply Lemma~\ref{lem:bin}. Again, we need to first make sure that the conditions of Lemma~\ref{lem:sr} and Lemma~\ref{lem:bin} are satisfied. As in the proof of Theorem~\ref{thm:iext}, these conditions are satisfied if the following conditions are satisfied.

\[k \geq 24 k^{3\alpha+\beta}\log n \text{ and } \ell=k^{\beta} \geq Ck^{\alpha}\log n\]
for some constant $C>1$.

Thus when $k \geq \frac{1}{2}\log^{2+\eta} n$, $\alpha=\frac{\mu}{6(2+\mu)}$, $\beta=\frac{6+2\mu}{6(2+\mu)}$, and $\mu=0.95\eta$, we have that for sufficiently large $n$,

\[\frac{k}{k^{3\alpha+\beta}}=k^{\frac{6+\mu}{6(2+\mu)}} > \Omega(\log^{1+\frac{\mu}{6}} n) > 24 \log n \text{ and } \frac{k^{\beta}}{k^{\alpha}}=k^{\frac{6+\mu}{6(2+\mu)}} \geq \Omega(\log^{1+\frac{\mu}{6}} n) > C \log n.\]

Also note that $\ell \leq k/(2h)$ thus the output length in each iteration of the lightest bin protocol also satisfies the condition of Lemma~\ref{lem:bin}. Thus we can apply that lemma. Note that we can first fix $Y_1$, and conditioned on this fixing $Z_1$ is a deterministic function of $X_1$, and thus at the end of the first iteration of the lightest bin protocol, we can use $Z_1$ to extract $Z_2$ from $Y_2$. By  Lemma~\ref{lem:bin}, again with probability $1-2^{-k^{\Omega(1)}}$ over the further fixing of $X_1$ (and thus $Z_1$), we will have that $Z_2$ has the $h$-wise independent property. Moreover now $Z_2$ is a  deterministic function of $Y_2$ and thus at the end of the next iteration of the lightest bin protocol, we can use $Z_2$ to extract $Z_3$ from $X_2$. Thus, since in the algorithm we are applying the lightest bin protocol in an ``alternating" manner, the whole algorithm works through as if we are dealing with independent sources. 

Note that the lightest bin protocol stops only if the number of rows in $Z_t$ is at most $\frac{16h^3}{\gamma^2}$. Thus before the iteration stops, we always have $N_t >\frac{16h^3}{\gamma^2}>h^3>h^2$. By Lemma~\ref{lem:bin} the probability of the ``bad event" in each iteration is at most $N_t^{-\sqrt{h}/2} < (h^3)^{-\sqrt{h}/2}=2^{-k^{\Omega(1)}}$. We now compute the number of iterations needed to decrease the number of rows from $N_1=\poly(n)$ to $\frac{16h^3}{\gamma^2}$.

In each iteration the number of rows in $Z_t$ decreases from $N_t$ to $N_{t+1} \leq \frac{16h}{\gamma^2} N_t^{\frac{2}{\sqrt{h}}}$. When $N_t \geq h^{\sqrt{h}}$, we have that $N_t^{\frac{2}{\sqrt{h}}} \geq h^2>\frac{16h}{\gamma^2}$. Thus 

\[N_{t+1} \leq \frac{16h}{\gamma^2} N_t^{\frac{2}{\sqrt{h}}}< N_t^{\frac{4}{\sqrt{h}}}.\]

Therefore, as long as $N_t \geq h^{\sqrt{h}}$, in each iteration the number of rows in $Z^t$ decreases from $N_t$ to $N_{t+1} \leq N_t^{\frac{4}{\sqrt{h}}}$. Since initially we have $N_1=\poly(n)$, the number of iterations needed to decrease the number of rows from $N_1=\poly(n)$ to $h^{\sqrt{h}}$ is at most $c'$ which equals

\begin{align*}
\log_{\frac{\sqrt{h}}{4}} \frac{\log N_1}{\sqrt{h} \log h} &=\frac{\log \log N_1-\frac{1}{2}\log h-\log \log h}{\frac{1}{2}\log h-2} \\
&= \frac{\log \log n+O(1)-\frac{1}{2}\log h-\log \log h}{\frac{1}{2}\log h-2} \leq \frac{\log \log n}{\frac{1}{2}\log h-2}-1 \\
& \leq \frac{\log \log n}{\frac{1}{2.1}\log h}-1 \leq \frac{\log \log n}{\frac{\alpha(2+\eta)}{2.2}\log \log n}-1 \mbox{  (since $k \geq \frac{1}{2}\log^{2+\eta} n$)}\\
& < \frac{13.2(2+\mu)}{\mu(2+\eta)}-1 <\frac{13.2}{\mu}-1 < \frac{14}{\eta}-1.
\end{align*}


Once $N_t \leq h^{\sqrt{h}}$, in the next iteration we have 

\[N_{t+1} \leq \frac{16h}{\gamma^2} N_t^{\frac{2}{\sqrt{h}}} \leq \frac{16h^3}{\gamma^2}.\]

Thus the number of iterations needed to decrease the number of rows from $N=\poly(n)$ to $\frac{16h^3}{\gamma^2}$ is at most $c_3=c'+1 < \frac{14}{\eta}$, which is also a constant. Since $\gamma=\frac{\eta}{70} < \frac{1}{5c_3}$, we have that in each $Z_t$, the fraction of ``good rows" is at least $\frac{2}{3}(1-\gamma)^{c_3}> \frac{2}{3}(1-c_3\gamma) \geq \frac{4}{5} \cdot \frac{2}{3} > 1/2$, which satisfies the requirement of  Lemma~\ref{lem:bin}. Also note that there exists a constant $C_0=C_0(\eta)$ such that whenever $n \geq C_0$ and $k \geq \log^2 n$ we have $h \geq k^{\alpha} \geq C_1$ where $C_1$ is the constant in Lemma~\ref{lem:bin}. Thus we are all good. Note that the number of blocks from $(X, Y)$ used in the iteration is at most $c_3+2$.

Finally, when we stop at step $t$, we can fix all previous blocks of $(X, Y)$ used in the algorithm except $(X', Y')$. Since the number of blocks is a constant, with probability $1-2^{-k^{\Omega(1)}}$ over this fixing, we have that $Z_{t-1}$ has the $h$-wise independent property as in Lemma~\ref{lem:bin}. Moreover now $Z_{t-1}$ is a deterministic function of $X$. Let $Z'_{t-1}$ be the concatenation of the rows of $Z_{t-1}$ with index in the output of the last lightest bin protocol. Note that $Z'_{t-1}$ is a deterministic function of $Z_{t-1}$ and has at most $\frac{16h^3}{\gamma^2}$ rows. Without loss of generality, we can assume that $Z'_{t-1}$ has exactly $\lfloor \frac{16h^3}{\gamma^2} \rfloor$ rows, otherwise we can add rows of all $0$ strings to it until this is achieved. This ensures that $Z'_{t-1}$ is a deterministic function of $Z_{t-1}$ with a fixed output domain. Thus the size of $Z'_{t-1}$ is bounded by $\frac{16h^3}{\gamma^2} \ell = O(k^{3\alpha+\beta})=o(k^{\frac{1+\mu}{2+\mu}})$. 

We now fix $Z'_{t-1}$. By Lemma~\ref{lem:condition} with probability $1-2^{-k^{\Omega(1)}}$ over the fixing of $Z'_{t-1}$, we have that $X'$ still has min-entropy $2k-o(k^{\frac{1+\mu}{2+\mu}})-k^{\Omega(1)}=2k-o(k)$. Also, by Lemma~\ref{lem:bin} with probability $1-2^{-k^{\Omega(1)}}$ over the fixing of $Z_{t-1}$ (and thus also the fixing of $Z'_{t-1}$), we have that $Z_t$ has the the $h$-wise independence property. Thus with probability $1-2^{-k^{\Omega(1)}}$ over the fixing of $Z'_{t-1}$, we have that $Z_t$ is $2^{-k^{\Omega(1)}}$-close to an SR-source.

Moreover, conditioned on the fixing of $Z'_{t-1}$, we have that $Z_t$ is a deterministic function of $Y'$, and is thus independent of $X'$. Thus by Theorem~\ref{thm:optext}, with probability $1-2^{-k^{\Omega(1)}}$ over the fixing of $Z_t$, we have that $Z'$ is $2^{-k^{\Omega(1)}}$-close to an SR-source (where each row has $1.9k$ bits). Note that the size of $Z_t$ is also bounded by $\frac{16h^3}{\gamma^2} \ell = O(k^{3\alpha+\beta})=o(k^{\frac{1+\mu}{2+\mu}})$. Thus again by Lemma~\ref{lem:condition} with probability $1-2^{-k^{\Omega(1)}}$ over the fixing of $Z_t$, we have that $Y'$ still has min-entropy $2k-o(k^{\frac{1+\mu}{2+\mu}})-k^{\Omega(1)}=2k-o(k)$. Moreover, conditioned on the fixing of $Z_t$, we have that $Z'$ is a deterministic function of $X'$, and is thus independent of $Y'$. Note that the number of rows in $Z'$ is at most $\frac{16h^3}{\gamma^2} =O(k^{3\alpha})= O(k^{\frac{3\mu}{6(2+\mu)}})$ and $k^{1-2 \cdot \frac{3\mu}{6(2+\mu)}}=k^{\frac{2}{2+\mu}}>\log^2 n$, thus by Theorem~\ref{thm:srgeneral} we have that 

\[\left |(W, Y')-(U_m, Y') \right | \leq 2^{-k^{\Omega(1)}}\]

and

\[\left |(W, Z')-(U_m, Z') \right | \leq 2^{-k^{\Omega(1)}},\]
where $m=1.8k$. Note that we have fixed all previously used blocks of $(X, Y)$, and now $Z'$ is a deterministic function of $X'$. Thus conditioned on the fixing of $Z'$, we have that $W$ is a deterministic function of $Y'$, and is thus independent of $X$. Therefore by adding back all the errors we also have

\[\left |(W, Y)-(U_m, Y) \right | \leq 2^{-k^{\Omega(1)}}\]

and

\[\left |(W, X)-(U_m, X) \right | \leq 2^{-k^{\Omega(1)}}.\]

Finally, note that the number of blocks required in each block source is at most $ \lceil \frac{c_3+2}{2} \rceil=\lceil \frac{7}{\eta} \rceil+1$.
\end{thmproof}

Note that when $n < C_0$, the extractor can be constructed in constant time just by exhaustive search (in fact, we can get a two-source extractor in this way). Thus, we have the following theorem (by replacing $2k$ with $k)$.

\BT
For every constant $\eta>0$ and all $n, k \in \N$ with $k \geq \log^{2+\eta} n$, there is an efficiently computable extractor $\bext: (\bits^n)^t \times (\bits^n)^t \to \bits^m$ with $t=\lceil \frac{7}{\eta} \rceil+1$, such that if $X=(X_1, X_2, \cdots X_t), Y=(Y_1, Y_2, \cdots Y_t)$ are two independent $(k ,k, \cdots, k)$- block sources where each block has $n$ bits, then

\[\left |(\bext(X, Y), Y)-(U_m, Y) \right | \leq \e\]

and

\[\left |(\bext(X, Y), X)-(U_m, X) \right | \leq \e,\]
where $m=0.9k$ and $\e=2^{-k^{\Omega(1)}}$. \footnote{The constant $0.9$ can be replaced by any constant less than $1$.}
\ET

As a corollary, we immediately obtain the following theorem. 

\BT
For every constant $\eta>0$ and all $n, k \in \N$ with $k \geq \log^{2+\eta} n$, there is an efficiently computable extractor $\iext: (\bits^n)^t \to \bits^m$ with $t=\lceil \frac{14}{\eta} \rceil+2$ such that if $X_1, \cdots, X_t$ are $t$ independent $(n,k)$-sources, then

\[\left |\iext(X_1, \cdots, X_t)-U_m \right | \leq \e,\]
where $m=0.9k$ and $\e=2^{-k^{\Omega(1)}}$. 
\ET


\section{Conclusions and Open Problems}\label{sec:conc}
In this paper we constructed an explicit extractor for three independent $(n, k)$ sources with min-entropy $k \geq \log^{12} n$, and error $\e=2^{-k^{\Omega(1)}}$. In fact our extractor works for one $(n, k)$ source and another independent $(k, k)$ block source. This improves the previously best known construction for general $(n, k)$ sources in \cite{Li13b}, and brings the construction of independent source extractors to nearly optimal. We also have improved results for the case of $k \geq \log^{2+\eta} n$ for any constant $\eta>0$, where we achieve a better constant-source extractor and in fact an extractor for two independent block sources with each having a constant number of blocks. As a by-product, we developed a general method to reduce the error in somewhere random sources from $1/\poly(n)$ to $2^{-\Omega(k)}$ while keeping the number of rows to be $\poly(n)$, at the cost of one extra weak source. 

Our new results essentially subsume all previous results about independent source extractors, except in the case of two-source extractors. The natural next step is thus to try to break the entropy rate $0.49$ barrier in Bourgain's extractor \cite{Bourgain05}. Another interesting direction is to use our techniques to build better two-source dispersers and Ramsey graphs, in the spirit of \cite{BarakRSW06}. Finally, it would be interesting to see if the techniques developed recently by the author in \cite{Li13a, Li13b} and here can be applied to the constructions of extractors and dispersers for other classes of sources, such as affine sources and small space sources.

\bibliographystyle{alpha}

\bibliography{refs}

\end{document}